\documentstyle[pra,aps,epsf]{revtex}

\newcommand{\ket}[1]{| #1 \rangle}
\newcommand{\bra}[1]{\langle #1 |}
\newcommand{\braket}[2]{\langle #1 | #2 \rangle}

\begin{document}
\draft
\preprint{sepnloc17 6/15/98}
\title{Quantum Nonlocality without Entanglement}

\author{Charles~H.~Bennett$^1$, David~P.~DiVincenzo$^1$, Christopher A.
Fuchs$^2$, Tal Mor$^3$, Eric Rains$^4$, Peter~W.~Shor$^4$,
John A. Smolin$^1$, and William K. Wootters$^5$}

\address{$^1$IBM Research Division, T. J. Watson Research Center,
Yorktown Heights, NY 10598} 
\address{$^2$Norman Bridge Laboratory of Physics 12-33, California
Institute of Technology, Pasadena, CA 91125}
\address{$^3$D\'epartement d'Informatique et de Recherche
Op\'erationelle, CP 6128, Succ. Centre-Ville, Montr\'eal H3C 3J7,
Canada}
\address{$^4$AT\&T Shannon Laboratory, 180 Park Ave. Bldg. 103, Florham
Park, NJ 07932}
\address{$^5$Physics Department, Williams College, Williamstown, MA
01267}

\date{\today}
\maketitle
\begin{abstract}

We exhibit an orthogonal set of product states of two three-state
particles that nevertheless cannot be reliably distinguished by a pair
of separated observers ignorant of which of the states has been
presented to them, even if the observers are allowed any sequence of
local operations and classical communication between the separate
observers.  It is proved that there is a finite gap between the mutual
information obtainable by a joint measurement on these states and a
measurement in which only local actions are permitted.  This result
implies the existence of separable superoperators that cannot be
implemented locally.  A set of states are found involving three
two-state particles which also appear to be nonmeasurable locally.
These and other multipartite states are classified according to the
entropy and entanglement costs of preparing and measuring them by
local operations.

\end{abstract}

\narrowtext
\widetext

\section{Introduction}

The most celebrated manifestations of quantum nonlocality arise from
entangled states---states of a compound quantum system that admit no
description in terms of states of the constituent parts. Entangled
states, by their experimentally confirmed violations of Bell-type
inequalities, provide strong evidence for the validity of quantum
mechanics, and they can be used for novel forms of information
processing, such as quantum cryptography~\cite{ekert},
entanglement-assisted communication~\cite{bfs,cleve98}, and quantum
teleportation~\cite{bbcjpw}, and for fast quantum
computations\cite{PS,Grover}, which pass through entangled states on
their way from a classical input to a classical output. A related
feature of quantum mechanics, also giving rise to nonclassical
behavior, is the impossibility of cloning\cite{WZ} or reliably
distinguishing nonorthogonal states.  Quantum systems that for one
reason or another behave classically (e.g., because they are of
macroscopic size or are coupled to a decohering environment) can
generally be described in terms of a set of orthogonal, unentangled
states.

In view of this, one might expect that if the states of a quantum
system were limited to a set of orthogonal product states, the system
would behave entirely classically, and would not exhibit any
nonlocality. In particular, if a compound quantum system, consisting
of two parts $A$ and $B$ held by separated observers (Alice and Bob),
were prepared by another party in one of several mutually orthogonal,
unentangled states, $\psi_1, \psi_2...\psi_n$ unknown to Alice and
Bob, then it ought to be possible to reliably discover which state the
system was in by locally measuring the separate parts.  Also, it ought
to be possible to clone the state of the whole by separately
duplicating the state of each part. We show that this is not the case,
by exhibiting sets of orthogonal, unentangled states $\{\psi_i\}$ of
two-party and three-party systems such that
\begin{itemize} \item the states $\{\psi_i\}$ can be reliably
distinguished by a joint measurement on the entire system, but not by
any sequence of local measurements on the parts, even with the help of
classical communication between the observers holding the separate
parts; \item the cloning operation
$\psi_i\rightarrow\psi_i\otimes\psi_i$ cannot be implemented by any
sequence of local operations and classical communication.\end{itemize}

Some of the features of this new kind of nonlocality appeared first in
\cite{PV}, which presented a set of orthogonal states of a bipartite
system that cannot be cloned if Alice and Bob cannot communicate at
all.  However, the states in \cite{PV} can be cloned if Alice and Bob
use one-way classical communication.

Many more of the nonlocal properties considered in the present work
were anticipated by the measurement protocol introduced by Peres and
Wootters\cite{PW}.  Their construction indicates the existence of a
nonlocality dual to that manifested by entangled systems: entangled
states must be prepared jointly, but exhibit anomalous correlations
when measured separately; the Peres-Wootters states are unentangled,
and can be prepared separately, but exhibit anomalous properties when
measured jointly.  We note that such anomalies are at the heart of
recent constructions for attaining the highest possible capacity of a
quantum channel for the transmission of classical
data\cite{Holevo79,Holevo96,Schumacher97a,Haus}.

In the Peres-Wootters scheme, the preparator chooses one of three
linear polarization directions 0, 60, or 120 degrees, and gives Alice
and Bob each one photon polarized in that direction. Their task is to
determine which of the three polarizations they have been given by a
sequence of separate measurements on the two photons, assisted by
classical communication between them, but they are not allowed to
perform joint measurements, nor to share entanglement, nor to exchange
quantum information.

Of course, because the three two-photon states are nonorthogonal, they
cannot be cloned or reliably distinguished, even by a joint
measurement. However, Peres and Wootters performed numerical
calculations which provided evidence (more evidence on an analogous
problem was provided by the work of Massar and Popescu\cite{MP})
indicating that a single joint measurement on both particles yielded
more information about the states than any sequence of local
measurements.  Thus unentangled nonorthogonal states appear to exhibit
a kind of quantitative nonlocality in their degree of
distinguishibility. The discovery of quantum teleportation,
incidentally, grew out of an attempt to identify what other resource,
besides actually being in the same place, would enable Alice and Bob
to make an optimal measurement of the Peres-Wootters states.

Another antecedent of the present work is a series of
papers\cite{GV,Asher,Tal98} resulting in the conclusion\cite{Tal98}
that several forms of quantum key distribution\cite{BB84} can be
viewed as involving orthogonal states of a serially-presented
bipartite system.  These states cannot be reliably distinguished by an
eavesdropper because she must let go of the first half of the system
before she receives the second half. In this example, the serial
time-ordering is essential: if, for example, the two parts were placed
in the hands of two separate classically-communicating eavesdroppers,
rather than being serially presented to one eavesdropper, the
eavesdroppers could easily cooperate to identify the state and break
the cryptosystem.

In this paper we report a form of nonlocality qualitatively stronger
than either of these antecedents. We extensively analyze an example in
which Alice and Bob are each given a three-state particle, and their
goal is to distinguish which of nine product states,
$\psi_i=\ket{\alpha_i}\otimes\ket{\beta_i}, i=1\ldots9$ the composite
$3\times 3$ quantum system was prepared in. Unlike the Peres-Wootters
example, these states are {\em orthogonal}, so the joint state could
be identified with perfect reliability by a collective measurement on
both particles.  However, the nine states are not orthogonal as seen
by Alice or Bob alone, and we prove that they cannot be reliably
distinguished by any sequence of local measurements, even permitting
an arbitrary amount of classical communication between Alice and
Bob. We call such a set of states ``locally immeasurable'' and give
other examples, e.g., a set of two mixed states of two two-state
particles (qubits), and sets of four or eight pure states of three
qubits, which apparently cannot be reliably distinguished by any local
procedure despite being orthogonal and unentangled.

In what sense is a locally immeasurable set of states ``nonlocal?''
Surely not in the usual sense of exhibiting phenomena inexplicable by
any local hidden variable (LHV) model. Because the $\psi_i$ are all
product states, it suffices to take the local states $\alpha_i$ and
$\beta_i$, on Alice's and Bob's side respectively, as the local hidden
variables.  The standard laws of quantum mechanics (e.g. Malus' law),
applied separately to Alice's and Bob's subsystems, can then explain
any local measurement statistics that may be observed.  However, an
essential feature of classical mechanics, not usually mentioned in LHV
discussions, is the fact that variables corresponding to real physical
properties are {\em not\/} hidden, but in principle measurable.  In
other words, classical mechanical systems admit a description in terms
of local {\em unhidden\/} variables.  The locally immeasurable sets of
quantum states we describe here are nonlocal in the sense that, if we
believe quantum mechanics, there is no local unhidden variable model
of their behavior.  Thus a measurement of the whole can reveal more
information about the system's state than any sequence of classically
coordinated measurements of the parts.

The inverse of local measurement is local preparation, the mapping
from a classically-provided index $i$ to the designated state
$\psi_i$, by local operations and classical communication.  If the
states $\psi_i$ are unentangled, local preparation is always possible,
but for any locally-immeasurable set of states this preparation
process is necessarily {\em irreversible\/} in the thermodynamic
sense, i.e., possible only when accompanied by a flow of entropy into
the environment. Of course if quantum communication or global
operations were allowed during preparation, the preparation could be
done reversibly, provided that the states being prepared are
orthogonal.

By eliminating certain states from a locally-immeasurable set (such as
$\{\psi_1,...\psi_9\}$ in Eq. (\ref{9states}) below), we obtain what
appears to be a weaker kind of nonlocality, in which the remaining
subset of states is both locally preparable and locally measurable,
but in neither case (so far as we have been able to discover) by a
thermodynamically reversible process.  Curiously, in these situations,
the entropy of preparation (by the best protocols we have been able to
find) exceeds the entropy of measurement.

Besides entropies of preparation and measurement we have explored
other quantitative measures of nonlocality for unentangled states.
One obvious measure is the amount of quantum communication that would
be needed to render an otherwise local measurement process reliable.
Another is the mutual information deficit when one attempts to
distinguish the states by the best local protocol.  Finally one can
quantify the amount of advice, from a third party who knows $i$, that
would be sufficient to guide Alice and Bob through an otherwise local
measurement procedure.

The results of this paper also have a bearing on, and were directly
motivated by, a question which arose recently in the context of a
different problem in quantum information processing.  This is the
problem of {\em entanglement purification}, in which Alice and Bob
have a large collection of identical bipartite mixed states that are
partially entangled.  Their object is to perform a sequence of
operations locally, i.e., by doing quantum operations on their halves
of the states and communicating classically, and end up with a smaller
number of pure, maximally entangled states.  Recently, bounds on the
efficiency of this process have been studied by Rains\cite{Rains} and
by Vedral and Plenio\cite{VP}; other constraints on entanglement
purification by separable superoperators have recently been studied
by Horodecki {\em et al.}\cite{Horod}.

In their work, they represent the sequence of operations using the
theory of {\em superoperators}, which can describe any combination of
unitary operations, interactions with an ancillary quantum system or
with the environment, quantum measurement, classical communication,
and subsequent quantum operations conditioned on measurement results.
In the operator-sum representation of superoperators developed by
Kraus and others, the general final state $\cal{S}(\rho)$
of the density operator of the system is written as a function of the
initial state $\rho$ as:
\begin{equation}
{\cal S}(\rho)=\sum_kS_k\rho S^\dagger_k.
\label{eq1}
\end{equation}
The $S_k$ operators appearing in this equation will be referred to as
``operation elements.''  A {\em trace-decreasing} superoperator
satisfies the condition $0\leq\sum_kS_k^\dagger S_k<1$ and is
appropriate for describing the effect of arbitrary quantum
measurements on the system (\cite{Bar}, Sec. III), while a {\em
trace-preserving} superoperator specified by $\sum_kS_k^\dagger S_k=1$
describes a general time evolution of the density operator if a
measurement is not made or its outcomes are ignored\cite{supops}.
Reference \cite{Niels} has a useful general review of the
superoperator formalism.

To impose the constraint that Alice and Bob act only locally, Rains,
and Vedral and Plenio, restricted their attention to {\em separable}
superoperators, in which the operation elements have a direct product
form involving an Alice operation and a Bob operation:
\begin{equation}
S_k=A_k\otimes B_k.
\label{eq2}
\end{equation}
We will show in Sec. \ref{lames} (see also \cite{Bar}, Sec. IX.C) that
all operations that Alice and Bob can perform during entanglement
purification {\em bilocally}, in which they can perform local quantum
operations and communicate classically, can be written in this
separable form.  This was enough for the derivation of valid upper
bounds on the efficiency of entanglement purification.  But the
natural question which this led to is the converse, that is, can all
separable superoperators be implemented by bilocal operations?

The answer to this question is definitely {\em no}, as a result of the
examples which we analyze in this paper.  Quantum measurements are a
subset of the superoperators, and measurements involving only product
states are separable superoperators.  Thus, our proof that some
unentangled states cannot be distinguished locally shows that some
separable superoperators cannot be implemented by only separate
operations by Alice and Bob with classical communication between them.
This indicates that any further investigations of entanglement
purification protocols involving separable superoperators will have to
be performed with some caution.

This paper is organized as follows: Section \ref{sec2} presents the
$3\times 3$ example and sketches the proof that these states cannot be
distinguished by local measurements.  Appendix \ref{appaa} gives many
of the important details of this proof, and Appendix \ref{peter}
supplies a crucial technical detail, that all superoperators can be
decomposed into a sequence of very weak measurements.  Section
\ref{sec3} shows how the measurement can be done locally if some
states are excluded, and presents the best measurement strategy we
have found for distinguishing (imperfectly) all nine states.  Section
\ref{sec4} shows how the measurement can be done for the $3\times 3$
example if entanglement is supplied.  Section \ref{sec5} analyzes the
thermodynamics of local state measurement, studying the heat generated
in measurement and in state preparation; Appendix \ref{appd} gives
some details.  Section \ref{tals} analyzes a three-party $2\times
2\times 2$ example involving 8 pure states.  Section \ref{disc} gives
other compact examples (4 pure states in a $2\times 2\times 2$ system,
2 mixed states in a $2\times 2$ system) and poses some questions for
the future (Appendix \ref{appc} gives details of a specific problem
considered there).

\section{A separable measurement which is not bilocal}
\label{sec2}

\subsection{The ensemble of states in a 3$\times$3 Hilbert space}

\begin{figure}[htbp]
\epsfxsize=8cm
\epsfbox{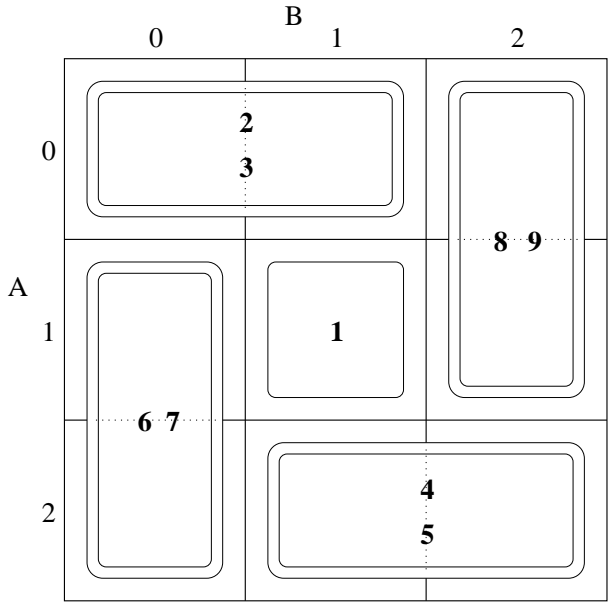}
\caption{A graphical depiction of the nine orthogonal states of Eq.
(\protect\ref{9states}) as a set of dominoes.}
\label{fig1}
\end{figure}

We will consider the following complete, orthonormal set of product
states $\psi_i=|\alpha_i\rangle\otimes|\beta_i\rangle$.
They live in a nine-dimensional Hilbert space, with Alice and
Bob each possessing three dimensions.  We will use the notation
$|0\rangle$, $|1\rangle$, and $|2\rangle$ for the bases of Alice's and
Bob's Hilbert spaces.  The orthonormal set is
\begin{equation}
\begin{array}{lll}
\,&|\alpha\rangle{\mbox \small (Alice)}\ &|\beta\rangle
{\mbox \small (Bob)}\\
\psi_1=&|1\rangle&|1\rangle\\
\psi_2=&|0\rangle&|0+1\rangle\\
\psi_3=&|0\rangle&|0-1\rangle\\
\psi_4=&|2\rangle&|1+2\rangle\\
\psi_5=&|2\rangle&|1-2\rangle\\
\psi_6=&|1+2\rangle&|0\rangle\\
\psi_7=&|1-2\rangle&|0\rangle\\
\psi_8=&|0+1\rangle&|2\rangle\\
\psi_9=&|0-1\rangle&|2\rangle.
\end{array}
\label{9states}
\end{equation}
Here $|0\pm 1\rangle$ stands for ${1\over\sqrt{2}}(|0\rangle\pm
|1\rangle)$, etc.  Figure \ref{fig1} shows a suggestive graphical way
to depict the 9 states of Eq. (\ref{9states}) in the $3\times 3$
Hilbert space of Alice and Bob.  The four dominoes represent the
four pairs of states that involve superpositions of the basis states.
State $\psi_1$ is clearly special in that it involves no such
superposition.

\subsection{The measurement}
\label{lames}

We will show that the separable superoperator ${\cal S}(\rho)=
\sum_iS_i\rho S_i^\dagger$ consisting of the projection operators
\begin{equation}
S_i=|i\rangle_A|i\rangle_B\langle\psi_i|
\label{supop}
\end{equation}
cannot be performed by local operations of Alice and Bob, even
allowing any amount of classical communication between them.  In
Eq. (\ref{supop}), the output Hilbert space is different from the
input; it is a space in which both Alice and Bob separately have a
complete and identical record of the outcome of the measurement.
See Sec. \ref{disc} for a discussion of why we use the particular
form of Eq. (\ref{supop}) for the operator; note that the input
state need not be present at the output in Eq. (\ref{supop}).

Since this superoperator corresponds to a standard von Neumann
measurement, we can equally well consider the problem in the form of
the following game: Alice and Bob are presented with one of the nine
orthonormal product states (for the time being, with equal prior
probabilities, let us say---this is not important, it is only
important that the prior probabilities of states $\psi_2$ through
$\psi_9$ be nonzero).  Their job is to agree on a measurement protocol
with which they can determine, with vanishingly small error, which of
the nine states it is, adhering to a bilocal protocol.

Let us characterize bilocal protocols a little more explicitly.  Our
discussion will apply both to bilocal measurements and to bilocal
superoperators (in which the measurement outcomes may be traced out).
By prior agreement one of the parties, let us say Alice, initiates the
sequence of operations.  The most general operation that she can
perform locally is specified by the set of operation elements
\begin{equation}
A_{r1}\otimes I.
\end{equation}
We will immediately specialize to the case where each value $r1$
labels a distinct ``round 1'' measurement outcome which she will
report to Bob, since no protocol in which she withheld any of this
information from Bob could have greater power.  She cannot act on
Bob's state, so her operators are always the identity $I$ on his
Hilbert space.  $A_{r1}$ can also include any unitary operation that
Alice may perform before or after the measurement.  Note also that the
operator $A_{r1}$ may not be a square matrix; the final Hilbert space
dimension may be smaller (but this would never be useful) or larger
(because of the introduction of an ancilla) than the original.

After the record $r1$ is reported to Bob, he does his own operation
\begin{equation}
I\otimes B_{r2}(r1).
\end{equation}
The only change from round 1 is that Bob's operations can be explicit
functions of the measurements reported in that round.  Now, the
process is repeated.  The overall set of operation elements specifying
the net operation after $n$ rounds is given by multiplying out
a sequence of these operations:
\begin{eqnarray}
&&S_m=A_m\otimes B_m,\label{2sep1}\\
&&A_m=A_{rn}(r1,r2,...,r(n-1))...A_{r3}(r1,r2)A_{r1},\label{2sep2}\\
&&B_m=B_{r(n-1)}(r1,r2,...,r(n-2))...B_{r4}(r1,r2,r3)B_{r2}(r1).
\label{2sep3}
\end{eqnarray}
Here the label $m$ can be thought of as a concatenation of all the
data collected through the $n$ rounds of measurement:
\begin{equation}
m=r1:r2:r3:...:rn.\label{concat}
\end{equation}
Equations (\ref{2sep1}-\ref{2sep3}) demonstrate the fact that all
bilocal operations are also separable operations.  It is the converse
statement that we are about to disprove for the operator corresponding
to the nine-state measurement, Eq. (\ref{supop}).

We can get some intuitive idea of why it will be hard for Alice and
Bob to perform Eq. (\ref{supop}) by local operations by noting the
result if Alice and Bob perform simple, local von Neumann measurements
in any of their rounds.  These measurements can be represented on the
``tic-tac-toe'' board of Fig. \ref{fig1} as simple horizontal or
vertical subdivisions of the board.  The fact that any such
subdivision cuts apart one of the dominoes shows very graphically that
after such an operation the distinguishability of the states is
spoiled.  This spoiling occurs in any local bases, and is more
formally just a reflection of the fact that the ensemble of states as
seen by Alice alone, or by Bob alone, is nonorthogonal.

However, it is not sufficient to show the impossibility of performing
Eq. (\ref{supop}) using a succession of local von Neumann
measurements, as Alice and Bob have available to them an infinite set
of weak measurement strategies\cite{Aharonov}.  Much more careful
reasoning is required to rule out any such strategy.  In the remainder
of this section we present the details of this proof, which also
results in a computation of an upper bound on the amount of
information Alice can Bob can obtain when attempting to perform the
nine-state measurement bilocally.

\subsection{Summary of the proof}

We assume that Alice and Bob have settled on a bilocal protocol with
which they will attempt to complete the measurement as well as
possible.  We identify the moment in the execution of this measurement
when Alice and Bob have accumulated a specific amount of partial
information.  We will have to show that it is always possible to
identify this moment, either in Alice and Bob's protocol or in an
equivalent protocol which can always be derived from theirs.  We then
show, based on the specific structure of the nine states, that at this
moment the nine possible input states must have become nonorthogonal
by a finite amount.  We then present an information-theoretic analysis
of the mutual information obtainable in the complete measurement, and
show, using an accessible-information bound, that the mutual
information obtainable by Alice and Bob two-locally is less, by a
finite amount, than the information obtained from a completely
nonlocal measurement.

Now we present the steps of this proof in detail.

\subsection{Information accumulation and the modified continuous
protocol}
\label{continn}

If the measurement has proceeded to a point where measurement record
$m$ has been obtained, an inference can be made using Bayes' theorem
of the probability $p(\psi_i|m)$ that the input state was
$\psi_i$:
\begin{equation}
p(\psi_i|m)={{p(m|\psi_i)p(\psi_i)}\over{\sum_j
p(m|\psi_j)p(\psi_j)}}
\label{bys}
\end{equation}
We take all prior probabilities $p(\psi_i)$ to be equal to $1\over 9$,
so they will drop out of this equation.  The measurement probabilities
$p(m|\psi_i)$ are given by the standard formula
\begin{equation}
p(m|\psi_i)={\rm Tr}(S_m|\psi_i\rangle\langle\psi_i|S^\dagger
_m)=\langle\psi_i|S^\dagger_mS_m|\psi_i\rangle.\label{stan}
\end{equation}
Here $S_m$ is the operation element of Eq. (\ref{2sep1}); the quantum
state in Alice's and Bob's possession has been transformed to
\begin{equation}
\phi_{i,m}\equiv S_m|\psi_i\rangle.\label{residuary}
\end{equation}

We imagine monitoring these prior probabilities every time a new round
is added to the measurement record in Eq. (\ref{concat}).  We will
divide the entire measurement into two stages, I and II; ``stage I''
of the measurement is declared to be complete when $p(\psi_i|m)$, for
some $i$, equals a particular value (the choice of this value is
discussed in detail in the next subsection).  ``Stage II'' is defined
as the entire operation from the end of stage I to the completion of
the protocol.

There is a problem with this, however: the measurement record changes
by discrete amounts on each round, and it is quite possible for these
probabilities to jump discontinuously when a new datum is appended to
this measurement record of Eq. (\ref{concat}).  Thus, it is likely
that the probabilities $p(\psi_i|m)$ will never attain any particular
value, but will jump past it at some particular round.  The
probabilities would evolve continuously only if Alice and Bob agree on
a protocol involving only weak measurements, for which all the
$A_{rk}$ and $B_{rk}$ of Eqs. (\ref{2sep2},\ref{2sep3}) are
approximately proportional to the identity operator.  But, in an
attempt to thwart the proof about to be given, Alice and Bob may agree
on a protocol which has both weak measurements and strong measurements
(for which the operators of Eqs. (\ref{2sep2},\ref{2sep3}) are not
approximately proportional to the identity).

However, such a strategy will never be helpful for Alice and Bob,
because, for any bilocal measurement protocol which they formulate
involving any combination of weak and strong measurements, a modified
measurement protocol exists that involves {\em only} weak measurements
for which the amount of information extracted by the overall
measurement is exactly the same.  For this modified protocol an
appropriate completion point for ``stage I'' of the measurement can
always be identified.  Thus we can prove, by the steps described
below, that the modified protocol cannot be completed successfully by
bilocal operations, and we give a bound on the attainable mutual
information of such a measurement.  But, since the modified protocol
is constructed to have the same measurement fidelity as the original
one, this proves that {\em any} protocol, involving any combination of
weak and strong measurements, also cannot attain perfect measurement
fidelity.

The modified protocol is created in a very simple way: it proceeds
through exactly the same steps as the original protocol, except that
at the point where the result of a strong measurement is about to be
reported to the other party by transmission through the classical
channel, the strong measurement record, treated as a
quantum-mechanical object, is itself subjected to a long sequence of
very weak measurements.  The outcomes of these weak measurements are
reported, one at a time, to the other party and appended to the
measurement record in Eq. (\ref{concat}).

The precise construction of this weak-measurement sequence is
described in Appendix \ref{peter}.  The weak measurements are designed
so that in their entirety they give almost perfect information about
the outcome of the strong measurement (the strong measurement outcome
itself can be reported at the end of this sequence as a confirmation).
So, the recipient of this steam of reports from the outcomes of the
weak measurements need only wait until they are done to know the
actual (strong) measurement outcome in order to proceed with the next
step of the original protocol.  But, except in cases with vanishingly
small probability, the information contained in the accumulating
measurement record grows continuously.

To conclude this discussion of the modified measurement protocol, we
can show how Alice and Bob can be duped into being unwitting
participants in the modified protocol, and also give an illuminating
if colloquial view of how the ``continuumization'' of the measurement
can take place.  What is required is a modification of the makeup of
the classical channel between Alice and Bob.  We imagine that when
Alice transmits the results of a measurement, thinking that it is
going directly into the classical channel to Bob, it is actually
intercepted by another party (Alice'), who performs the necessary
sequence of weak measurements.  Here is a way that Alice' can
implement this operation:  She examines the bit transmitted by Alice.
If the bit is a 0, she selects a slightly head-biassed coin, flips it
many times, each time transmitting the outcome into the classical
channel.  If the bit is a 1, she does the same thing with a slightly
tail-biassed coin.  At the other end of the channel there is another
intercepting agent (Bob') who, after studying a long enough string of
coin flips sent by Alice', can with high confidence deduce the coin
bias and report the result to Bob.  Alice and Bob are oblivious to
this whole intervening process; nevertheless, as measured by the data
actually passing through the channel, the modified protocol with
nearly continuous evolution of the available information has been
achieved.

\subsection{The state of affairs after stage I of the measurement}

Having established that no matter what Alice and Bob's measurement
protocol, we can view the probabilities as evolving continuously
in time, we can declare that stage I of the measurement is complete
when
\begin{equation}
\max_i\ p(\psi_i|m_I)={1\over 9}+\epsilon,
\label{depart}
\end{equation}
that is, after the probabilities have evolved by a small but finite
amount away from their initial value of $1\over 9$.  It should be
noted that since some measurement outcomes might be much more
informative than others, the time of completion of stage I is not
fixed; it will in general require a greater number of rounds for one
measurement record $m_I$ than for another.

The $\epsilon$ in Eq. (\ref{depart}) should be some definite, small,
but noninfinitesimal number.  Moreover, we will require that all
posterior probabilities $p(\psi_i|m_I)$ be nonzero.  For this any
value smaller than $1\over 72$ will be acceptable, since
\begin{equation}
\min_i\ p(\psi_i|m_I)\geq{1\over 9}-8\epsilon.
\label{smallest}
\end{equation}
We now rewrite Bayes' theorem from Eq. (\ref{bys}):
\begin{equation}
p(\psi_i|m_I)={{\langle\psi_i|E_{m_I}|\psi_i\rangle}\over
{\sum_j\langle\psi_j|E_{m_I}|\psi_j\rangle}}
={{\langle \alpha_i|a_{m_I}| \alpha_i\rangle
\langle \beta_i|b_{m_I}| \beta_i\rangle}\over
{\sum_j\langle\psi_j|E_{m_I}|\psi_j\rangle}}.\label{pform}
\end{equation}
Here we have introduced an abbreviated notation for several
operators which will come up repeatedly in the upcoming derivations:
\begin{equation}
\begin{array}{l}
E_{m_I}=S^\dagger_{m_I}S_{m_I}=a_{m_I}\otimes b_{m_I},\nonumber\\
a_{m_I}=A^\dagger_{m_I}A_{m_I},\\
b_{m_I}=B^\dagger_{m_I}B_{m_I}.\nonumber
\end{array}
\label{aandb}
\end{equation}
Where there is no risk of confusion we will drop the index $m_I$ from
$E_{m_I}$, $a_{m_I}$, and $b_{m_I}$.

It is easy to bound the greatest possible spread in the
probability distribution:
\begin{equation}
{{8+72\epsilon}\over{8-9\epsilon}}\leq\max_{i,j}{{p(\psi_i|m_I)}
\over{p(\psi_j|m_I)}}=\max_{i,j}
{{\langle \alpha_i|a|\alpha_i\rangle\langle \beta_i|b| \beta_i\rangle} \over
{\langle \alpha_j|a|\alpha_j\rangle\langle \beta_j|b|\beta_j\rangle}}
\leq{{1+9\epsilon}\over {1-72\epsilon}}.
\label{bound}
\end{equation}
An important technical consequence of declaring stage I complete at
this point is that it is guaranteed
that all the matrix elements $\langle \alpha_i|a|\alpha_i\rangle$ and
$\langle \beta_i|b|\beta_i\rangle$ are nonzero; this condition
will be used repeatedly in the analysis of Appendix \ref{appaa}
(to be described shortly).  The more crucial condition from
Eq. (\ref{bound}) is that either the following equation is true,
\begin{equation}
\max_{i,j}{{\langle \alpha_i|a|\alpha_i\rangle}
\over{\langle \alpha_j|a|\alpha_j\rangle}}\geq
\sqrt{{8+72\epsilon}\over{8-9\epsilon}},
\label{boundmod2}
\end{equation}
or the corresponding equation for $b$ is true.  This says that either
the operator $a$ or $b$ differs from being proportional to the
identity operator by a finite amount.  This will be the key fact in
the analysis we are about to report.

The basic idea is that at the completion of stage I, from Alice's and
Bob's point of view there is a nonzero probability that the initial
state was any one of the nine.  In order for Alice and Bob to complete
the job of identifying which state they have been given---with a
reliability approaching 100\%---it is necessary that the nine states
remaining after stage I, Eq.~(\ref{residuary}), still be almost
perfectly distinguishable.  That is, the states must still be nearly
{\em orthogonal}.  But we can show that, because of
Eq.~(\ref{boundmod2}), these residual states cannot be sufficiently
orthogonal to complete the task.  In fact, we will be able to compute
exactly to what extent they must be nonorthogonal.  For we can show
that, if we assume that the overlap of any two of these residual
states is $\delta$ or less, i.e.,
\begin{eqnarray}
\max_{i,j}\ \langle\phi_{i,m_I}|\phi_{j,m_I}\rangle=
\max_{i,j}{{|\langle\psi_i|a\otimes b|\psi_j\rangle|}\over
{\sqrt{\langle\psi_i|a\otimes b|\psi_i\rangle \langle\psi_j|a\otimes
b|\psi_j\rangle}}}=\delta,
\label{orcon}
\end{eqnarray}
then both $a$ and $b$ will both be almost proportional to the identity
operator, with relative corrections proportional to $\delta$.  This is
done in Appendix~\ref{appaa} where these corrections are derived
precisely.  The important consequence of this is that
\begin{equation}
\max_{i,j}{{\langle \alpha_i|a| \alpha_i\rangle}
\over{\langle \alpha_j|a| \alpha_j\rangle}}\leq 1+O(\delta),
\label{boundmod4}
\end{equation}
and the same for $b$.  Equations (\ref{boundmod2}) and
(\ref{boundmod4}) cannot be both satisfied unless
$\delta=O(\epsilon)$, that is, unless the residual states are
nonorthogonal by a finite amount.

So, at this point we can conclude that the measurement
Eq. (\ref{supop}) cannot be done bilocally, except with less that
100\% accuracy; this is the main result that we set out to prove.  We
now proceed to a more quantitative analysis of bilocal approximations
to this measurement.

\subsection{Information-theoretic analysis of the two-stage
measurement}

We can now perform an analysis of the precise effects of this
nonorthogonality, and derive an upper bound on the information
attainable by Alice and Bob from any bilocal protocol.  We will use
the standard classical quantifier of information, the {\em mutual
information}\cite{Abr}, which gives the amount of knowledge of one
random variable (in our case, the identity of quantum state
$\psi_i$) gained by having a knowledge of another (here, the
outcome of the measurement).

Recall that we have broken the measurement by Alice and Bob into two
stages.  We will call the random variable describing the stage-I
outcomes $M_I$.  The outcomes of all subsequent (stage-II)
measurements will be denoted by random variable $M_{II}$.  Alice and
Bob's object is to deduce perfectly the label $i$ of one of the nine
states $\psi_i$ (Eq. (\ref{9states})); we will use the symbol
$W$ for this random variable (for ``which wavefunction'').  We
quantify the information attainable in the measurement by the mutual
information $I(W;M_I,M_{II})$ between $W$ and the composite
measurement outcomes $M_I$ and $M_{II}$.  For a perfect measurement,
the attainable mutual information is $\log_29$; we will show that
$I(W;M_I,M_{II})$ must be less than this.  We first use the additivity
property of mutual information (\cite{Abr}, p. 125) to write:
\begin{equation}
I(W;M_I,M_{II})=I(W;M_{II}|M_I)+I(W;M_I).\label{mi1}
\end{equation}
This expression introduces the mutual information between $W$ and
$M_{II}$ conditional on $M_I$, which can be written as an average
over all the possible outcomes $m_I$ of the measurement in stage I:
\begin{equation}
I(W;M_{II}|M_I)=\sum_{m_I}p(m_I)I(W;M_{II}|m_I).\label{mi2}
\end{equation}
Now, combining Eqs. (\ref{mi1},\ref{mi2}) with the definition of the
mutual information:
\begin{equation}
I(W;M_I)=H(W)-H(W|M_I),\label{mi22}
\end{equation}
and using the fact that the entropy of the initial distribution
$H(W)=\log_29$, we obtain
\begin{equation}
I(W;M_I,M_{II})=\log_29-\sum_{m_I}p(m_I)(H(W|m_I)-I(W;M_{II}|m_I)).
\label{mi3}
\end{equation}
To show that Eq. (\ref{mi3}) must be less than $\log_29$ it will be
sufficient to show that each member of the sum is strictly positive.
The conditions at the end stage I make it possible for us to do
this.

To make things explicit, let us suppose that at the end of stage I the
residual quantum states (recall Eq. (\ref{residuary}))
$\rho_i=|\phi_{i,m_I}\rangle\langle\phi_{i,m_I}|$ occur with
probabilities $q_i=p(\psi_i|m_I)$ from Eq. (\ref{pform}).  (There will
be no confusion from leaving out the $m_I$ label.)  Moreover, let us
suppose that the measurement to be performed in stage II corresponds
to a positive operator-valued measure $\{M_b\}$ fixed by measurement
outcome $m_I$.  Then the explicit expression for the mutual
information $I(W;M_{II}|m_I)=I(M_{II};W|m_I)$ becomes
\begin{eqnarray}
I(M_{II};W|m_I)&=&H(M_{II}|m_I)-H(M_{II}|W,m_I),\nonumber\\
&=&-\sum_b({\rm tr}\,\rho M_b)\log_2({\rm tr}\,\rho M_b)
+\sum_{i=1}^9 q_i
\sum_b({\rm tr}\,\rho_i M_b)\log_2({\rm tr}\,\rho_i M_b)\;,
\end{eqnarray}
where $\rho=\sum_i q_i\rho_i$.  Note that
$H(W|m_I)=-\sum_{i=1}^9q_i\log_2q_i$.

Without loss of generality for the present set of manipulations, let
us take $\phi_{1,m_I}$ and $\phi_{2,m_I}$ to be the two states assured
to have a nonvanishing overlap
$\langle\phi_{1,m_I}|\phi_{2,m_I}\rangle=\delta$ (recall
Eq. (\ref{orcon})).  We may partition the density operator $\rho$
according to the two states that interest us most as follows.  Let
\begin{equation}
\tau_1=\sum_{i=1}^2 \frac{q_i}{s_1}\rho_i
\qquad\mbox{and}\qquad
\tau_2=\sum_{i=3}^9 \frac{q_i}{s_2}\rho_i
\label{Joplin}
\end{equation}
where $s_1=q_1+q_2$ and $s_2=1-s_1$.  We can think of this partition
as generating two new ``which wavefunction'' random variables $W_1$
and $W_2$---the probabilities associated with these random variables
are just the renormalized ones appearing in Eq.~(\ref{Joplin}).  Note
that $\rho=s_1\tau_1+s_2\tau_2$.  Then, by the classic converse to the
concavity of the Shannon entropy (\cite{OhyaPetz}, p.~21), it follows
that
\begin{equation}
-\sum_b({\rm tr}\,\rho M_b)\log_2({\rm tr}\,\rho M_b)\le
-s_1\sum_b({\rm tr}\,\tau_1 M_b)\log_2({\rm tr}\,\tau_1 M_b)
-s_2\sum_b({\rm tr}\,\tau_2 M_b)\log_2({\rm tr}\,\tau_2 M_b)+
h(s_1)\;,
\end{equation}
where $h(x)=-x\log_2 x-(1-x)\log_2(1-x)$ is the binary entropy
function.
Hence, if we write
\begin{eqnarray}
I(M_{II};W_1|m_I)&=&-\sum_b({\rm tr}\,\tau_1 M_b)
\log_2({\rm tr}\,\tau_1 M_b)+
\sum_{i=1}^2 \frac{q_i}{s_1}
\sum_b({\rm tr}\,\rho_i M_b)\log_2({\rm tr}\,\rho_i M_b)\;,
\label{Chris1}
\\
I(M_{II};W_2|m_I)&=&-\sum_b({\rm tr}\,\tau_2 M_b)
\log_2({\rm tr}\,\tau_2 M_b)+
\sum_{i=3}^9 \frac{q_i}{s_2}
\sum_b({\rm tr}\,\rho_i M_b)\log_2({\rm tr}\,\rho_i M_b)\;,
\end{eqnarray}
it follows that
\begin{equation}
I(M_{II};W|m_I)\le s_1 I(M_{II};W_1|m_I) + s_2 I(M_{II};W_2|m_I)
+h(s_1)\;.\label{CF1}
\end{equation}
We can further bound this, so as to remove all dependence on states
$\phi_{3,m_I}$ through $\phi_{9,m_I}$, by noting that
\begin{equation}
I(M_{II};W_2|m_I)\le H(W_2|m_I)=-\sum_{i=3}^9\frac{q_i}{s_2}\log_2
\frac{q_i}{s_2}\;.\label{CF2}
\end{equation}
Combining Eqs. (\ref{CF1}) and (\ref{CF2}) gives
\begin{equation}
H(W|m_I)-I(W;M_{II}|m_I)\ge-\sum_{i=1}^2 q_i\log_2 q_i + s_1\log_2
s_1-s_1 I(M_{II};W_1|m_I)\;.
\label{DeanFlorentino}
\end{equation}

Equation~(\ref{DeanFlorentino}) can be further bounded so as to remove
any explicit dependence on $q_1$ and $q_2$ by noting that, for fixed
$s_1$, the first term in the expression on the right-hand side is
minimized when $q_1=q_2$.  (One can verify this simply by taking a
derivative respect to one of the free variables.)  Making that
restriction, one can see furthermore that the resultant term is
monotonically increasing in $q_1$.  Thus the bound we are looking for
can be found by taking $q_1$ to be its minimal allowed value, namely
$q_1=\beta={1\over 9}-8\epsilon$ (recall Eq. (\ref{smallest})).  With
all that in place, we have that
\begin{equation}
H(W|m_I)-I(W;M_{II}|m_I)\ge2\beta\!\left[1+
\sum_b({\rm tr}\,\tau_1 M_b)\log_2({\rm tr}\,\tau_1 M_b)-
\sum_{i=1}^2 \frac{1}{2}
\sum_b({\rm tr}\,\rho_i M_b)\log_2({\rm tr}\,\rho_i M_b)\right]\;,
\label{DisBeLloydBaby}
\end{equation}
where now $\tau_1=\frac{1}{2}(\rho_1+\rho_2)$.

Finally it is a question of removing all dependence on the quantum
measurement $\{M_b\}$.  This can be gotten by noting that the two
right-most terms in the right-hand side Eq.~(\ref{DisBeLloydBaby})
simply correspond to the mutual information given by the measurement
$\{M_b\}$ about the two equiprobable nonorthogonal quantum states
$\phi_{1,m_I}$ and $\phi_{2,m_I}$ (cf. Eq. (\ref{Chris1})).
Optimizing over all quantum measurements, we obtain the accessible
information of those two states \cite{LevitinFuchs}.  Inserting that
into Eq.~(\ref{DisBeLloydBaby}) and recalling Eq. (\ref{orcon}) we
finally find,
\begin{eqnarray}
&H(W|m_I)-I(W;M_{II}|m_I)\geq 2\beta\, h({1\over 2}-{1\over
2}\sqrt{1-\delta^2})=({2\over 9}-16\epsilon) h({1\over 2}-{1\over
2}\sqrt{1-\delta^2}),&
\label{PSbound}
\end{eqnarray}
where $h(x)$ is again the binary entropy.

The last bound can be made useful by establishing a quantitative link
between $\epsilon$ and $\delta$ in Eq. (\ref{PSbound}).  To do this,
we must identify the value of $\delta$ for which, given all the
constraints derived in Appendix~\ref{appaa}, it is first possible to
satisfy Eq.~(\ref{boundmod2}) for some values of $i$ and $j$.  It is
this value of $\delta$ which must be used in the bound
Eq.~(\ref{PSbound}).  We have exhaustively examined all $i,j$ pairs to
determine which one allows the greatest ratio of $a$ (or $b$) matrix
elements for a given value of $\delta$.  We find this to be the case
for $i=8$ and $j=6$ in Eq.~(\ref{9states}) (or other
symmetry-equivalent ones).  For this choice we can write
\begin{equation}
{{\langle x_8|a|x_8\rangle}\over{\langle x_6|a|x_6\rangle}}=
{{a_{00}+a_{11}+2{\rm Re}\,a_{01}}\over
{a_{11}+a_{22}+2{\rm Re}\,a_{12}}}.
\end{equation}
This ratio attains its maximum value when
\begin{equation}
a_{00}=a_{11}\frac{1+\delta}{1-\delta}\;,\quad
a_{22}=a_{11}\frac{1-\delta}{1+\delta}\;,\quad
{\rm Re}\,a_{01}=a_{11}\nu_\epsilon\sqrt{\frac{1+\delta}{1-\delta}}
\;,\quad
{\rm Re}\,a_{12}=-a_{11}\nu_\epsilon\sqrt{\frac{1-\delta}{1+\delta}}
\end{equation}
These are the extremal values permitted by Eqs.~(\ref{AsiaMinor})
and (\ref{thebiggie}).  The value this gives is
\begin{equation}
\max_{i,j}{{\langle \alpha_i|a| \alpha_i\rangle}
\over{\langle \alpha_j|a| \alpha_j\rangle}}
\leq f_\epsilon(\delta)=\left(
{{1+\delta}\over{1-\delta}}\right){{1+\nu_\epsilon\sqrt{1-\delta^2}}
\over{1-\nu_\epsilon\sqrt{1-\delta^2}}}.
\label{boundmod3}
\end{equation}
The smallest value of $\delta$ for which Eqs. (\ref{boundmod2}) and
(\ref{boundmod3}) are consistent is given by the solution to the
equation
\begin{equation}
f_\epsilon(\delta)=\sqrt{{8+72\epsilon}\over{8-9\epsilon}}.
\label{plugin}
\end{equation}
Using Mathematica, we have found the choice of $\epsilon$ and $\delta$
consistent with Eq. (\ref{plugin}) that gives the strongest bound on
the mutual information in Eq. (\ref{PSbound}).  We obtain:
\begin{eqnarray}
&I(W;M_I,M_{II})\leq\log_29-\Delta,&\label{fin}
\end{eqnarray}
where the mutual-information deficit $\Delta=0.00000531$.  This upper
bound is attained when $\epsilon=0.00823$, corresponding to a
nonorthogonality parameter $\delta=0.00344$ and a minimum-probability
parameter $\beta=0.0453=(0.408)/9$.  Thus, we bound the information
attainable by bilocal operations by Alice and Bob away from that
attainable in a fully nonlocal measurement by a minute but finite
amount.

\section{Searching for Optimal Local Measurements}
\label{sec3}

Equation~(\ref{fin}) gives our upper bound on the mutual information
one can obtain by means of local operations and classical
communication.  However, it is unlikely that this bound is a close
approximation to the actual optimal mutual information accessible in
this way; most likely the optimal value is significantly lower.  In
this section we explore specific measurement strategies for our
nine-state ensemble in order to get a sense of how well one can in
fact distinguish the states by local means.  We will thereby obtain a
{\em lower} bound on the mutual information.

We begin by considering a simpler problem, namely, distinguishing
only eight of the nine states from each other.  That is, we consider the
case where the prior probability of one of the states is zero.

As we noted earlier, state $\psi_1$ from Eq. (\ref{9states}) is
special.  In fact, it is never used in the analysis of Appendix
\ref{appaa}; thus, its presence or absence is irrelevant to the
nonorthogonality conditions which we have derived.  This means that
this state is not necessary to make the measurement undoable
bilocally.  Thus, even if we take the prior probabilities of the
states such that $p(\psi_1)=0$, we will still reach the conclusion
that the full mutual information is unattainable by a bilocal
procedure (the quantitative analysis will be different than that given
above).

\begin{figure}[htbp]
\epsfxsize=8cm
\epsfbox{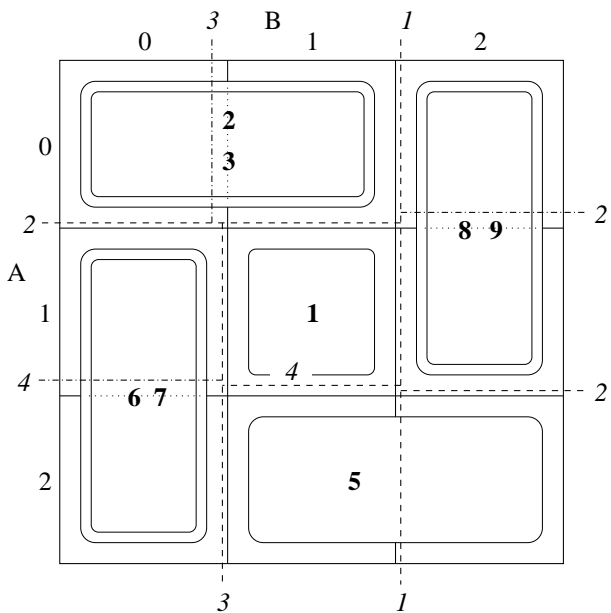}
\caption{The sequence of measurements performed to distinguish the
states of Fig. \protect\ref{fig1} if state $\psi_4$ is excluded.
The dashed lines indicate the von Neumann measurements, the italic
numbers indicate the order in which they are performed.  Dash-dotted
lines indicate measurements in the rotated basis.}
\label{fig2}
\end{figure}

\begin{figure}[htbp]
\epsfxsize=8cm
\epsfbox{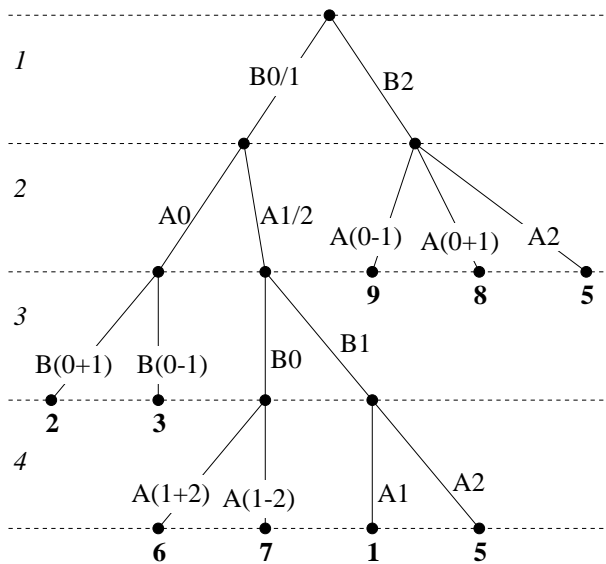}
\caption{A tree depicting the four states of measurement indicated in
Fig. \protect\ref{fig2}.  ``A'' and ``B'' indicate the party
performing the measurement.  B0/1 indicates that the 0 and 1 outcomes
are not distinguished.  The bold-faced numbers at the base of the tree
indicate the states which are inferred from this chain of
measurements.}
\label{fig3}
\end{figure}

The same is not true for the other states: if the prior probability of
any of the states $\psi_2$-$\psi_9$ is zero, then the measurement can
be completed successfully by Alice and Bob.  Figures \ref{fig2} and
\ref{fig3} illustrate this for the case when the state $\psi_4$ is
left out.  One way of explaining the strategy is that since the 4-5
domino of Fig. \ref{fig2} is no longer complete, it can be cut by
a von Neumann measurement, which will disturb state
$\psi_5$ but still leave
it distinguishable from all the other eight states.  Thus, the
protocol can begin with cut 1 of Fig. \ref{fig2}, which corresponds to
an incomplete von-Neumann measurement by Bob which distinguishes his
state $|2\rangle$ from states $|0\rangle$ or $|1\rangle$ (but does not
distinguish between $|0\rangle$ and $|1\rangle$).  The next step to be
taken by Alice depends on the reported outcome as received by her from
Bob, as indicated by the tree of Fig. \ref{fig3}; likewise all four
rounds of the measurement are similarly contingent on the measurement
outcomes of preceding rounds.  The object at every round is to move
towards isolating a domino so that its pair of states can be
distinguished by a measurement in the rotated basis.

We now turn to our original problem of distinguishing optimally among
all nine states, assumed to have equal prior probabilities.  The
measurement strategy just described is a reasonable one to pursue even
when all nine states are present.  It accurately distinguishes states
$\psi_{1-3}$ and $\psi_{6-9}$, and it distinguishes these states from
$\psi_4$ and $\psi_5$; it fails only to distinguish these last two
states from each other.  (In applying Fig. \ref{fig3} to this case,
one should imagine replacing ``5'' with ``4 or 5.'')  Thus if Alice
and Bob use this measurement, then with probability $\frac{7}{9}$ they
obtain the full $\log_2 9$ bits of information, and with probability
$\frac{2}{9}$ they are left one bit short; so the mutual information
is $\log_2 9 - \frac{2}{9}$ = 2.9477 bits.  One can, however, do
better, and we now present a series of improvements over the above
strategy.

We may express the improved measurements as sequences of positive
operator-valued measures (POVMs).  For example, Bob could start with a
POVM consisting of elements $b_{r1}$ (these are 3$\times$3 matrices
that must satisfy the constraint $\sum_{r1}b_{r1} = I$), after which
Alice will perform a measurement $\{a_{r2}\}$, and so on.  As it
happens, all of our improved measurements can be represented in terms
of POVMs whose elements are diagonal in the standard bases for Alice
and Bob.  It is therefore convenient to represent these POVM elements
by their diagonal values.  For example, in the measurement described
above, Bob's opening POVM (in this case a von Neumann measurement),
which distinguishes his state $|2\rangle$ from $|0\rangle$ and
$|1\rangle$, has two elements which we represent as $\{1,1,0\}$ and
$\{0,0,1\}$.

Our first improvement is to replace this von Neumann measurement by a
more symmetric POVM whose elements are $b_1 = \{1,\frac{1}{2},0\}$ and
$b_2 = \{0,\frac{1}{2},1\}$.  (If Bob were to perform this measurement
when his part of the system was in the central state $|1\rangle$, the
outcome would be random.)  Note that each outcome of this measurement
rules out one of the columns of Fig. \ref{fig1}; that is, it rules out
one of Bob's states $|0\rangle$ or $|2\rangle$.  Once this has been
done, Alice may freely cut either the 6-7 domino or the 8-9 domino,
and from this point Bob and Alice may proceed as above to find out
(with no further damage) in which domino the actual state lies.
However, Bob's initial measurement damages both the 2-3 domino and the
4-5 domino, so that at the end, he will not be able to distinguish
perfectly between $\psi_2$ and $\psi_3$ or between $\psi_4$ and
$\psi_5$.  Thus, in order to evaluate the mutual information
obtainable via this strategy, we need to know the effect of Bob's
initial POVM on these four states.  This effect depends on what
operation element $B_{r1}$ we choose to associate with the POVM
element $b_{r1}$.  Any $B_{r1}$ satisfying
$B_{r1}^{\dag}B_{r1}=b_{r1}$ is allowed, but it is simplest to let
$B_{r1}$ be $|r1\rangle \otimes \sqrt{b_{r1}}$, where $|r1\rangle$ is
the classical record of the outcome.  To see how this measurement
affects the states, let us suppose that the actual state is $\psi_4$,
so that Bob's part of the system begins in the state
$|\phi\rangle=\frac{1}{\sqrt{2}}(|1\rangle + |2\rangle)$.  Then if Bob
gets the outcome $b_1$, the final state of Bob's part of the system
(not including the classical record) is
$\sqrt{b_1}|\phi\rangle=\frac{1}{2}|1\rangle$, and if he gets the
outcome $b_2$ the final state is $\sqrt{b_2}|\phi\rangle =
\frac{1}{2}|1\rangle + \frac{1}{\sqrt{2}}|2\rangle$.  (These states
are automatically subnormalized so that their squared norms are the
probabilities of the corresponding outcomes, namely, $\frac{1}{4}$ and
$\frac{3}{4}$.)  If the initial state had been $\psi_5$, then the
results would have been the same but with $|2\rangle$ replaced by
$-|2\rangle$.  Thus the first outcome renders $\psi_4$ and $\psi_5$
completely indistinguishable, while the second merely makes them
non-orthogonal.  In the latter case Bob can, at the end, try to
determine whether the original state was $\psi_4$ or $\psi_5$ by
performing the optimal measurement for distinguishing two equally
likely non-orthogonal states \cite{LevitinFuchs}.  In this case the
optimal measurement is simply the orthogonal measurement whose
outcomes are B($1+2$) and B($1-2$).  Similar considerations apply to
the states $\psi_2$ or $\psi_3$.  One finds that this strategy yields
a mutual information of 2.9964 bits, which is an improvement over the
strategy of Fig. \ref{fig3}.

A further improvement is gained by replacing Bob's initial POVM by a
less informative and less destructive one whose elements are
$\{p,\frac{1}{2},1-p\}$ and $\{1-p,\frac{1}{2},p\}$, where
$\frac{1}{2} < p < 1$.  The rest of the measurement is left unchanged.
Optimizing over $p$, one finds that this strategy can yield 3.009 bits
of mutual information.  Note, however, that in this case Bob's initial
measurement does not rule out any column of Figure 1, so that when
Alice later cuts a domino, she may be cutting the actual state, in
which case her action will cost them one bit.  One may suspect that
Alice should be more careful, and indeed the mutual information is
improved if she makes a weaker measurement.  In fact, the best
strategy we have found delays until the fourth round a measurement
that guarantees the complete cutting of a domino.

This best strategy consists of the following steps, in which the
values of the parameters $p,q,r,s,$ and $t$ are to be determined by
optimization:
\begin{enumerate}
\item Bob: $\{p,\frac{1}{2},1-p\}$ {\em vs} $\{1-p,\frac{1}{2},p\}$.
Let us assume that Bob gets the first outcome.  (In the other case all
the POVM elements appearing in the succeeding steps have their
diagonal values reversed; that is, the roles of states $|0\rangle$ and
$|2\rangle$ are interchanged.)
\item Alice: $\{0,1-q,1-r\}$ {\em vs} $\{1,q,r\}$.  The first outcome
cuts the 8-9 domino, and we go directly to step 5.  The second outcome
makes it safer for Bob to risk cutting the 4-5 domino, so we proceed
to step 3.
\item Bob: $\{1-s,1-t,0\}$ {\em vs} $\{s,t,1\}$.  The first outcome
cuts the 4-5 domino, and we go directly to step 5.  The second outcome
makes it safer for Alice to cut the 6-7 domino, so we proceed to step
4.
\item Alice: $\{1,1,0\}$ {\em vs} $\{0,0,1\}$.  Either outcome cuts
the 6-7 domino.
\item At this point some domino has been cut, so that Alice and Bob
can proceed as above to determine in which domino the actual state
lies.  If this domino contains two states that have not been collapsed
into the same state, Alice and Bob then perform a measurement to try
to distinguish them.
\end{enumerate}
Optimizing over the values of the parameters, we find that the mutual
information is $\log_2 9 - 0.1575 = 3.0125$ bits.  (One set of
parameter values giving this result is $p=0.726,\, q=0.395,\,
r=0.312,\, s=0.071,\, t=0.104$.)  Moreover, numerical evidence
indicates that no further advantage is gained by allowing another
round before making a firm cut (it would be a cut of the 2-3 domino,
as we proceed clockwise around the grid).  Thus it is conceivable that
this value of the mutual information is indeed optimal, though we
cannot rule out an entirely different strategy that does better.

Summarizing the results of this Section and the preceding one, we have
\begin{eqnarray}
&\log_2 9 - 0.1575 \le I(W; M_I, M_{II}) \le \log_2 9 - \Delta.&
\end{eqnarray}

Note that the results presented in this section can be seen as a
realization of the ideas behind our proof in Section \ref{sec2}.
Alice and Bob begin by performing a sequence of POVMs aimed at
determining in which domino the actual state lies; this sequence can
be thought of as stage I of the measurement.  At this point, just as
in our proof, the states remaining to be distinguished have become
non-orthogonal, so that the final mutual information must fall short
of $\log_2 9$ bits.

\begin{figure}[htbp]
\epsfxsize=8cm
\epsfbox{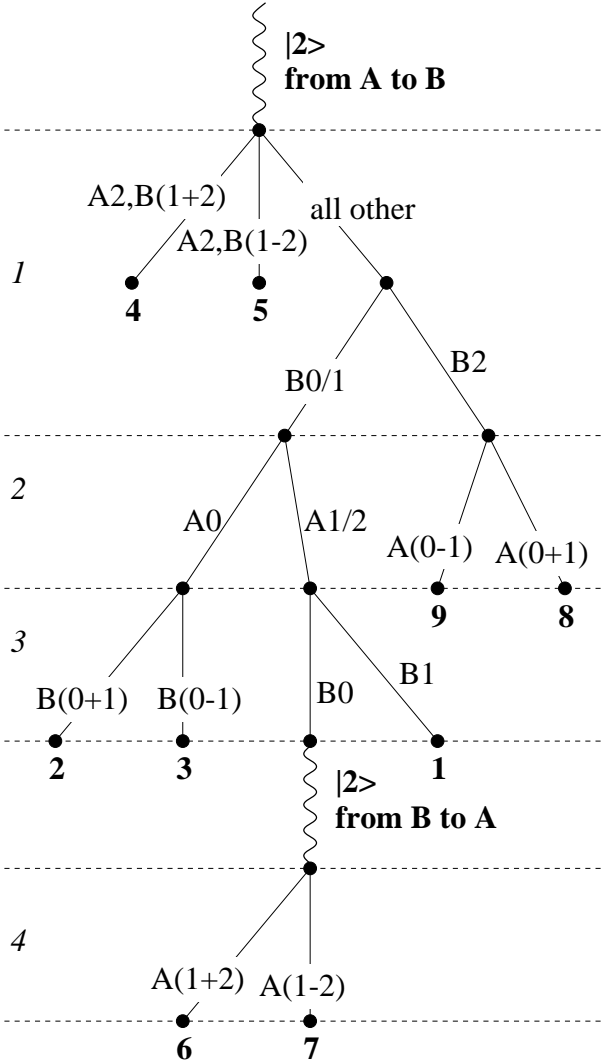}
\caption{A modification of the tree of Fig. \protect\ref{fig3} which
shows how all nine states can be reliably distinguished with some
quantum communication from Alice to Bob.  The wavy lines indicate
the episodes of quantum transmission; the first transmission permits
Bob to locally do a measurement involving both A and B pieces of the
Hilbert space.}
\label{fig4}
\end{figure}

\section{A realization of the two-party separable superoperator with
shared qubits}
\label{sec4}

Having established that the measurement can only be done approximately
if Alice and Bob only communicate classically, it is natural to ask
what quantum resources would permit them to complete the measurement.
It is obvious that they can do it if Alice ships her entire 3-state
system to Bob and he performs the full operation in his lab, reporting
the result classically back to Alice.  In the case of all 9 states
having equal prior probability, this requires the transmission of
$\log_23\approx 1.58496$ qubits.  If state $\psi_1$ is left out and
the other 8 states are equiprobable, the density matrix of the state
held by Alice has less than maximal entropy, in fact it has
$h_3({3\over 8},{2\over 8},{3\over 8})={11\over 4}-\log_23\approx
1.16504$ bits of entropy.  Using the Schumacher compression
theorem\cite{BS}, this means that if Alice and Bob are performing
many shots of the same measurement on states drawn from the same
ensemble, then the quantum transmission from Alice and Bob can be
compressed to $1.16504$ qubits per shot.

However, in the nine-state case we can exhibit a protocol for
completing the measurement which requires a smaller overall number of
qubits transmitted.  It starts with the imperfect protocol involving
only classical communication just discussed (Fig. \ref{fig4}), and
adds a part to permit states 4 and 5 to be perfectly distinguished.
This will require only $h({1\over 3})+{2\over 9}\approx 1.14152$
qubits (over many repetitions of the measurement).  For the 8-state
case the protocol will actually be worse than the straightforward one,
requiring $h({3\over 8})+{2\over 8}\approx 1.20443$ qubits of
transmission.  In neither case do we know that the procedures which we
discuss here are optimal.

The modified protocol for the 9-state case begins with Alice
transmitting the $|2\rangle$ component of her Hilbert space to Bob.
It is obvious that she could do this by sending 1 qubit, if she adopts
a 3-qubit unary encoding of her Hilbert space, i.e.,
$|0\rangle\rightarrow|100\rangle$, $|1\rangle\rightarrow|010\rangle$,
and $|2\rangle\rightarrow|001\rangle$.  In fact the third qubit in
this representation has less than maximal entropy, having entropy
$h({1\over 3})$ (it has higher entropy, $h({3\over 8})$, for the
8-state case).  Thus, again using Schumacher's theorem\cite{BS}, the
transmission can be compressed over many realizations of the
measurement so that only $h({1\over 3})$ of a qubit per measurement
needs to be transmitted.

As indicated by the tree in Fig. \ref{fig4}, Bob's possession of
$|2\rangle_A$ permits him to immediately do a measurement which
distinguishes whether the state is $\psi_4$, $\psi_5$, or is one of
the others.  After this has been done the sequence of measurements
proceeds identically as in the classical protocol (Fig. 4), except
that some possibilities can be pruned off as they correspond to
$\psi_4$ and $\psi_5$ cases which have already been distinguished.
Before completing round 4, Alice must be again in possession of
$|2\rangle_A$, which requires a qubit transmission back from Bob.
This qubit is not compressible, but this transmission will only be
required if the state is $\psi_6$ or $\psi_7$, which will only happen
$2\over 9$ of the time, and will count as $2\over 9$ qubits of
transmission ($2\over 8$ for the 8-state case).

Adding up the qubit transmissions at the beginning and the end of
Fig. \ref{fig4} gives $h({1\over 3})+{2\over 9}\approx 1.14152$ qubits
as mentioned above.  This transmission can be made unidirectional,
since a qubit sent in one direction, if it is entangled with a qubit
left behind, may always be used to teleport a qubit in the opposite
direction\cite{bbcjpw}.  Note that even with the assistance of qubit
transmissions, this protocol requires several rounds of classical
transmission; it is a true ``two-way'' protocol, that is, requiring
bidirectional classical communication\cite{BDSW}.

\section{Thermodynamics of nonlocal measurements and state
preparation}
\label{sec5}

\subsection{Irreversibility of measurement}

We now explore another information-theoretic feature of our two-party
measurement that illustrates in another way the nonlocality of this
orthogonal measurement. If the parts of the quantum states are
assembled in one location, then a measurement in any orthogonal basis,
in addition to being doable with 100\% fidelity, can be done {\em
reversibly}. That is, the quantum state can be converted into
classical data without any discarding of information to the
environment. Therefore by Landauer's principle~\cite{Landauer} no heat
is generated during the measurement.  The reversible method can be
illustrated by a simple qubit example: if the measurement is to
distinguish $\ket{0}$ from $\ket{1}$, and the classical record of the
bit is to be stored in the macro states
$\ket{\underline{0}}\equiv\ket{000...}$ and
$\ket{\underline{1}}\equiv\ket{111...}$ (containing, say, $10^{23}$
qubits), then the procedure involves starting the macro system in a
standard state (so that the initial states of the system to be
measured is either $\ket{0000...}$ or $\ket{1000...}$), then
performing repeated quantum XOR operations\cite{BDSW} with the qubit
to be measured as the source and all the qubits of the macro state as
the targets. In the end, the measured qubit may as well be considered
to be part of the macro system containing the classical answer. Note
that no interaction with any other environment is necessary to
complete this, or any other, local orthogonal measurement.

The situation is rather different for our two-party orthogonal
measurement. Suppose that we consider a case in which the measurement
can be achieved by Alice and Bob, for example the case in which state
$\psi_4$ is promised not to be present. Although Alice and Bob can
perform this measurement, they clearly cannot do so reversibly, i.e.,
as a finite sequence of local reversible operations and classical
communications. In the protocol described in Fig. \ref{fig2}, the
irreversibility arises in the first step, where, if the state is
$\psi_5$, it is irreversibly transformed to either state
$\ket{2}\ket{1}$ or to $\ket{2}\ket{2}$. Thus in this case 1 bit of
entropy is produced.  If each of the eight permitted states occurs
with equal probability, then the average entropy generated is $1\over
8$ of a bit. We cannot prove this entropy of measurement is minimal,
though we have found no more efficient protocol. Many other cases can
be easily worked out; for example, if it is promised that the state is
only one of four (say, $\psi_6$, $\psi_2$, $\psi_8$, and $\psi_4$),
then $1 \over 4$ of a bit of entropy will be generated by the obvious
protocol.

It appears that reversible measurements are only possible if the set
of states can be progressively dissected by Alice and Bob without
breaking any dominoes. To formalize this notion, we introduce a few
definitions.  Let $S=\{\psi_i\}$ be a set of pure product states
shared between Alice and Bob, where
$\psi_i=\alpha_i\otimes\beta_i$. Given such a set, we define a {\em
splitting\/} of $S$ by Alice as a partition of $S$ into two nonempty
disjoint subsets $S=S_1\cup S_2$ such that for all $\psi_i\in S_1$ and
for all $\psi_j$ in $S_2$, $\braket{\alpha_i}{\alpha_j}=0$. A
splitting by Bob is defined similarly. A set $S$ is {\em
dissectible\/} if there is a tree each of whose interior nodes is a
splitting by Alice or Bob and whose leaves are singletons. For
example, using the numbering of Eq.~(\ref{9states}) and
Fig.~\ref{fig1}, the set $\{\psi_2,\psi_6,\psi_8\}$ is dissectible,
but $\{\psi_2,\psi_4,\psi_6\,\psi_8\}$ is not. The dissectibility of
an arbitrary set $S$ can be determined by examining finitely many
possible splitting trees. Clearly any subset of a dissectible set is
dissectible. It is evident that if an ensemble of states ${\cal
E}=\{p_i,\psi_i\}$ has support only on a dissectible set, then both
its entropy of preparation and entropy of measurement are zero. It is
tempting to argue that, conversely, nondissectible sets, if they are
locally measurable at all, have positive entropies of measurement, but
to be sure of this, one would have to exclude the (unlikely-seeming)
possibility of multi-step measurement procedures which, while not
strictly reversible for any finite $n$, would succeed in identifying
each of the states in the nondissectible set with error probability and
entropy production both tending to 0 in the limit of large $n$.

A further analysis of this irreversibility reveals that it can be
thought of as originating in the necessity for classical communication
between Alice and Bob. In order to assure that the channel between
them can convey only classical and no quantum information, the channel
itself must possess a quantum environment (in order to dephase the
data passing though it). This raises the possibility that Alice or Bob
will be obliged to become entangled with the environment of the
channel in the course of communicating the necessary classical
information, thereby causing themselves to have a finite amount of
entropy. Exactly the same amount must also appear in the channel
environment. When, for example, Alice and Bob have been given state
$\psi_5=\ket{2}\otimes(\ket{1}+\ket{2})$, and Bob sends the result of
his first measurement in Fig. \ref{fig3} (collapsing his state to a
mixture of $\ket{1}$ and $\ket{2}$) to Alice, he has created
entanglement between the measurement outcome and the environment, so
that the joint system of message and environment is left an entangled
state of the form $\ket{1}\otimes e_1+\ket{2}\otimes e_2$, where $e_1$
and $e_2$ are two orthogonal states of the environment.

Note that measurement protocols requiring classical communication are
not inevitably irreversible. For example, for the dissectible set
$\{\psi_2,\psi_6,\psi_8\}$ considered previously, a bit of communication
from Bob to Alice is required to complete the measurement; still no
entropy is generated. This is so because this bit is guaranteed to be in
one of the computational basis states, precisely the states with which
the dephasing channel does not entangle. It is the necessity, in the
above example, of delivering a bit to the channel which is in a
superposition of basis states which leads to the entanglement and the
irreversibility.

\subsection{Irreversibility of state preparation}

For dissectible sets of states, such as $\{\psi_2,\psi_6,\psi_8\}$,
the mapping
\begin{equation}
\ket{i}\otimes\ket{i}\leftrightarrow\ket{\alpha_i}\otimes\ket{\beta_i},
\end{equation}
(using the notation of Eq. (\ref{9states})) between classical
instructions and the state described is locally reversible and can be
performed in either direction without the generation of waste
information. Conversely, nondissectible sets, such as
$\{\psi_2,\psi_4,\psi_6,\psi_8\}$, cannot be prepared by any finite
sequence of reversible operations, and we conjecture that even
asymptotic multistep protocols could not reduce either the heat of
preparation or the heat of measurement to zero. Perhaps surprisingly,
the heats of preparation and measurement, by the best protocols we
have been able to discover, are unequal.

\begin{figure}[htbp]
\epsfxsize=8cm
\epsfbox{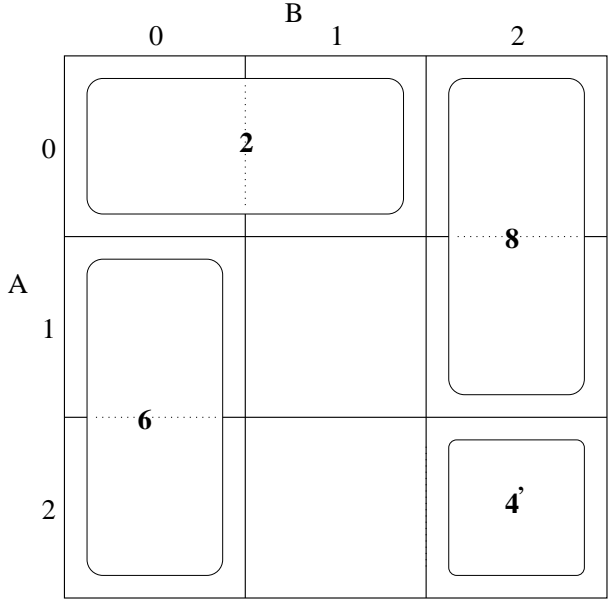}
\caption{A set of four states, shown in the domino notation, which
can be prepared locally by Alice and Bob in a reversible fashion.}
\label{fig8}
\end{figure}

To give an example of irreversible state preparation, consider the
following method for the preparation for the nondissectible set
$\{\psi_2, \psi_4, \psi_6, \psi_8\}$ mentioned above. The protocol,
which is the best we know, will produce $h({1\over 4})\approx 0.811$
bits of entropy, considerably more than the entropy of
measurement. The procedure works as follows: First, Bob computes a
function $f$ of the preparation instruction $i$ which records whether
the state to be synthesized is $\psi_4$ ($f(4)=1$) or one of the
others ($f(2,6,8)=0$), saving the result in a work bit. Then Alice and
Bob reversibly prepare the modified four states of Fig. \ref{fig8};
that is, if the instruction is to prepare $\psi_4$, $\psi_{4'}$ is
prepared, and in the other three cases exactly the desired state is
produced.

This preparation can be carried out reversibly because the modified
set $\{\psi_2, \psi_4', \psi_6, \psi_8\}$ is dissectible. Next, Bob
performs a Hadamard rotation on his state
($|2\rangle\rightarrow|1+2\rangle$, $|1\rangle\rightarrow|1-2\rangle$,
$|0\rangle\rightarrow|0\rangle$) conditional upon the state of $f(i)$,
which transforms $4'$ into $4$ and leaves the other three states
unchanged as desired. Finally Bob erases his work bit $f(i)$, which
requires discarding $h({1\over 4})$ bits of entropy into the
environment.  Similar reasoning shows that the equiprobable nine-state
ensemble can be prepared at a cost of $h({2\over 9})\approx0.764$, and
the equiprobable eight-state ensemble (without the center state) at a
cost of $h({2\over 8})\approx0.811$ bits of entropy.

It should perhaps be noted that the local preparation and measurement
protocols we have described, while irreversible from the viewpoint of
Alice and Bob, become reversible when viewed from a global perspective,
including Bob, Alice and the environment. In the preparation protocol we
have just described this global reversibility arises because the waste
classical information $f(i)$ discarded into the environment in the last
step is not random, but instead is entirely determined by the joint
state $\psi_i$ of Alice and Bob. Therefore discarding it, though it
increases the entropy of the environment, does not increase the entropy
of the universe. The global reversibility of the measurement protocol
for this same set of four states arises because the information
discarded into the environment in the final stage is merely the other
half of the entanglement created at an earlier stage of the protocol,
when one of the dominoes might have been collapsed. Thus the final act
of discarding restores the environment to a pure state.

When speaking of the thermodynamic costs of local preparation and
local measurement, it should be recalled that, although any set of
product states can be locally prepared, not all sets can be locally
measured.  The full set of nine states $\{\psi_1...\psi_9\}$ of
Eq.~(\ref{9states}), for example, is not locally measurable at all, no
matter how much heat generation is allowed. Conversely, there are sets
of pure bipartite states that cannot be {\em prepared\/} locally, even
with the generation of heat, because one or more states in the set is
entangled. The concepts of entropy of preparation and entropy of
measurement can nevertheless be extended to such sets, indeed to any
orthogonal set of pure bipartite states, by allowing Alice and Bob to
draw on a reservoir of prior entanglement (e.g., standard singlets
$\Psi^-={1\over\sqrt{2}}(\ket{01}-\ket{10})$ shared between them) to
help perform actions, such as teleportation\cite{bbcjpw}, that could
not otherwise be done locally. In this fashion one can define an
entanglement-assisted entropy of local preparation, and an
entanglement-assisted entropy of local measurement. In
entanglement-assisted measurement, an otherwise immeasurable set like
the original set of nine states is rendered measurable by teleporting
quantum information as required, say, in the protocol of Fig.
\ref{fig4}.  However, each teleportation generates two bits of waste
classical information per qubit teleported, thereby contributing to
the entropy of measurement. Again we can calculate the amounts of
entanglement consumed and entropy produced by simple protocols,
without knowing whether more efficient ones exist. The protocols
described earlier give an entanglement-assisted entropy of measurement
of 2.28304 bits for the equiprobable nine-state ensemble, and 2.40886
bits for the eight-state ensemble (omitting the central state), in
each case twice the amount of entanglement consumed, because the
protocols generate no other waste information aside from that
associated with the teleportations. Turning now to
entanglement-assisted preparation, a typical set of states requiring
entanglement to prepare from classical directions is set of four Bell
states\cite{BDSW} $\{\Phi^+,\Phi^-,\Psi^+,\Psi^-\}$. The entropy of
preparation by the obvious protocol in this case is two bits per state
prepared (Bob reads the classical directions, applies an appropriate
Pauli rotation to the standard $\Psi^-$ to make the desired Bell
state, then throws away the classical directions).

Finally suppose Alice and Bob are given an unknown member of the 9-state
set (or some other locally-immeasurable set) and wish to determine which
state they have without the help of entanglement, but with some hints
from a person who knows which state they have been given. We define the
``advice of measurement'' as the minimal amount of advice needed (in
conjunction with their own local actions) to guide Alice and Bob to the
right answer. As we have seen above, a negative hint like ``the state is
not $\psi_4$'' is sufficient. This might appear to be a lot of advice
(as much as a totally informative positive hint like ``the state is
$\psi_3$,'') but in fact such negative hints are highly compressed by
classical hashing techniques, asymptotically requiring only ${8\over
9}\log_2{8\over 7}\approx 0.171\,$ bits per hint in the nine-state
case.  Appendix \ref{appd} gives details of the compression of these 
types of hints.  

We note, however, that the non-von Neumann measurements discussed at
the end of section III allow an even more efficient form of advice.
There it was shown that an appropriate POVM yields
$3.0125=\log9-0.1575$ bits of information about the unknown state in
the 9-state case; therefore, after Alice and Bob have performed their
POVM, only 0.1575 bits of additional information need be provided
asymptotically for them to identify the state exactly.

As an aside, we note that the value of advice, and the amount needed,
may depend on its timing. Although in the 9-state measurement problem
the most efficient advice we know of can safely be given at the end,
after the POVM has been completed, there are other situations in
quantum information theory, not to mention in everyday life, when
early advice is more useful than late advice. In the BB84 quantum key
distribution protocol\cite{BB84}, for example, the basis information
may be regarded as a form of advice that is delayed to make it less
useful to the eavesdropper. In a deterministic setting, where the
adviser can foresee all future events, nothing is lost by giving all
necessary advice at the beginning. But when unforeseen events are
possible, the most efficient kind of advice, better than prior or
posterior advice, may be as-needed or concurrent advice. Suppose Alice
and Bob are about to begin a long car trip. They ask their more
experienced friend Eve which route to take. A few days later they
telephone again, asking her how to repair a flat tire. To be helpful,
the route advice must be given at the beginning, but it would be
wasteful to give the repair advice then because the flat tire might
not have happened. The prominent role of measurements, whether von
Neumann or POVM, with unpredictable outcomes, in our analysis of the
9-state problem suggests that as-needed advice might be the optimal
kind here also.

The notion of advice of measurement can be extended to sets of entangled
states as well, for example the set of four Bell states. Here one bit of
advice is sufficient (e.g., whether the unknown Bell state is of the $+$
or $-$ type), since the other bit ($\Phi$ vs. $\Psi$) can be learned by
comparing the results of local measurements in the $z$ basis. The
table summarizes the various measures of nonlocality for some
of the ensembles we have been considering.

\begin{table}
\begin{center}
\begin{tabular}{lll ccc}

Ensemble       \ \     & 9-state &  2468   &  246  & 4-Bell & 2-Bell \\
\hline
Locally Preparable \ \ &   Yes   &  Yes    &  Yes  &  No     &  No        \\
Locally Measurable \ \ &   No    &  Yes    &  Yes  &  No     &  Yes       \\
Dissectible        \ \ &   No    &  No     &  Yes  &  No     &  No        \\
Entropy of Prep.   \ \ & 0.764   &  0.811  &   0   &   2     &  1         \\
Entropy of Meas. \ \ & 2.283   &  0.250  &   0   &   2     &  1         \\
Entanglement of Prep.   \ \ &   0     &  0      &   0   &   1     &  1       \\
Entanglement of Meas. \ \ & 1.142   &  0      &   0   &   1     &  0         \\
Advice  of Meas. \ \ & 0.1575  &  0      &   0   &   1     &  0         \\
\hline
\end{tabular}\end{center}
{\small Notes: Entropies, entanglements, and advice for non-Bell
ensembles are upper bounds from known protocols---actual values could
be less.  Entropy of measurement for 9-state and 4-Bell ensembles are
for entanglement-assisted measurement, since these ensembles are
otherwise not locally measurable.  The nine-state ensemble consists of
9 equiprobable states $\psi_1...\psi_9$ of Eq.~(\ref{9states}) and
Fig.~\ref{fig1}.  The 2468 and 246 ensembles are equiprobable
distributions over $\{\psi_2,\psi_4,\psi_6,\psi_8\}$ and
$\{\psi_2,\psi_4,\psi_6\}$ respectively.  The 4-Bell ensemble consists
of four equiprobable Bell states $\{\Phi^+,\Phi^-,\Psi^+,\Psi^-\}$,
and the 2-Bell ensemble of two equiprobable Bell states, e.g.,
$\{\Phi^+,\Psi^+\}$}.
\end{table}

\section{A three-party separable superoperator}
\label{tals}

We shall now show another example of a separable von Neumann
measurement, this time involving three parties, Alice, Bob and Carol,
each holding just a qubit (two-state system).  While we have not
performed a full analysis of this case, it appears to have the same
properties as the 9-state measurement above---that partial measurement
causes indistinguishability of the residual states---suggesting that
this is another case in which the measurement cannot be done locally
by the three parties, even if the three can partake in any amount of
classical communication among themselves.  The superoperator involves
a complete orthonormal set of eight product states living in the
eight-dimensional Hilbert space.  This appears to be the smallest
possible Hilbert space which still presents such behavior (it is easy
to show, using a simple elimination process, that a qubit-trit system
or a qubit-qubit system is not sufficient).  The eight states are:
\begin{equation}
\begin{array}{llllr}
\,&{\mbox \small Alice}&{\mbox \small Bob}&{\mbox \small Carol}\\
\phi_1=&|0\rangle&|0\rangle&|0\rangle&\,\,\,\,\,=\ \verb%000%\\
\phi_2=&|1\rangle&|1\rangle&|1\rangle&=\ \verb%111%\\
\phi_3=&|0+1\rangle&|0\rangle&|1\rangle&=\ \verb%+01%\\
\phi_4=&|0-1\rangle&|0\rangle&|1\rangle&=\ \verb%-01%\\
\phi_5=&|0\rangle&|1\rangle&|0+1\rangle&=\ \verb%01+%\\
\phi_6=&|0\rangle&|1\rangle&|0-1\rangle&=\ \verb%01-%\\
\phi_7=&|1\rangle&|0+1\rangle&|0\rangle&=\ \verb%1+0%\\
\phi_8=&|1\rangle&|0-1\rangle&|0\rangle&=\ \verb%1-0%\end{array}
\label{8states}
\end{equation}
(leaving out normalizations).  On the right side of these equations we
introduce an obvious shorthand for these states which we will use in
the Discussion.  We will indicate the evidence that the separable
superoperator consisting of the projection operators
\begin{equation}
S_i=|i\rangle_A|i\rangle_B|i\rangle_C\langle\phi_i|
\label{supop3}
\end{equation}
cannot be performed by 3-local operations, in which Alice, Bob, and
Carol can only perform local quantum operations and broadcast
classical information to each other.

The arguments are equivalent to those in the two-trit example, and
again rely on considering any measurement as a two-stage process.  In
the case where all prior probabilities are equal ($1\over 8$ in this
case), we declare stage I to be complete when
\begin{equation}
\max_i p(\phi_i|m_I)={1\over 8}+\epsilon,
\end{equation}
with some positive $\epsilon$ smaller than $1\over 56$, It is again
simple to bound the greatest possible spread of the probability
distribution:
\begin{equation}
{{7+56\epsilon}\over{7-8\epsilon}}\leq\max_{i,j}{{p(\phi_i|m_I)}
\over{p(\phi_j|m_I)}}=\max_{i,j}{{\langle\phi_i|E|\phi_i
\rangle}\over{\langle\phi_j|E|\phi_j\rangle}}\leq{{1+8\epsilon}\over
{1-56\epsilon}}.
\label{bound3}
\end{equation}
As before, this equation guarantees that all diagonal matrix elements
of $E$, $\langle\phi_i|E|\phi_i \rangle=\langle\phi_i|a\otimes
b\otimes c|\phi_i \rangle$, are nonzero, and it also guarantees that
the maximum and minimum matrix elements are different.  Also as
before, we can show that the states after stage I become
nonorthogonal, which should permit us to derive a definite
mutual-information deficit.  We will not develop this proof here, but
we will give a simple sketch of how we prove that the states are
nonorthogonal.  We will just show here that the states cannot be {\em
exactly} orthogonal:
\begin{equation}
|\langle\phi_j|a\otimes b\otimes c|\phi_i\rangle|=0,
\;\;\;\;\forall i\neq j.
\label{orcon3}
\end{equation}
This proof can be generalized step by step into a full analysis as in
Appendix \ref{appaa}.

1) Writing the orthogonality condition for $i=3$ and $j=4$ gives
condition that
\begin{equation}
(a_{00}+a_{01}-a_{10}-a_{11})b_{00}c_{11}=0.
\end{equation}
Since diagonal matrix elements of $b$ and $c$ must be nonzero by
the arguments from Eq. (\ref{bound3}), the $a$ factor must be zero;
taking the real part gives
\begin{equation}
a_{00}=a_{11}.
\end{equation}

2) Taking taking $i=5$ and $j=6$ and applying
the same reasoning gives
\begin{equation}
c_{00}=c_{11}.
\end{equation}

3) And taking $i=7$ and $j=8$ gives
\begin{equation}
b_{00}=b_{11}.
\end{equation}

4) Now we write the four orthogonality conditions coming from all
combinations of $i=3,4$ and $j=5,6$:
\begin{equation}
\begin{array}{rcl}
(a_{00}+a_{01})b_{01}(c_{10}+c_{11})&=&0\\
-(a_{00}+a_{01})b_{01}(c_{10}-c_{11})&=&0\\
(a_{00}-a_{01})b_{01}(c_{10}+c_{11})&=&0\\
-(a_{00}-a_{01})b_{01}(c_{10}-c_{11})&=&0.\end{array}
\end{equation}
Adding these four equations gives
\begin{equation}
4a_{00}b_{01}c_{11}=0.
\end{equation}
Since $a_{00}\neq 0$ and $c_{11}\neq 0$, we conclude that
\begin{equation}
b_{01}=0=b_{10}.
\end{equation}

5) Doing the same for the equations involving $i=3,4$ and
$j=7,8$ gives
\begin{equation}
c_{01}=0=c_{10}.
\end{equation}

6) And finally from the equations involving $i=5,6$ and
$j=7,8$, we get
\begin{equation}
a_{01}=0=a_{10}.
\end{equation}

Putting observations 1) through 6) together, we conclude that $a$, $b$
and $c$ must be proportional to the identity operator.  But this is
inconsistent with Eq. (\ref{bound3}), which established that the
different diagonal matrix elements of $E$ must differ by a finite
amount.  When developed more fully, this result should contradict the
assumption that the measurement could be done even approximately by
3-local operations.

\begin{figure}[htbp]
\epsfxsize=8cm
\epsfbox{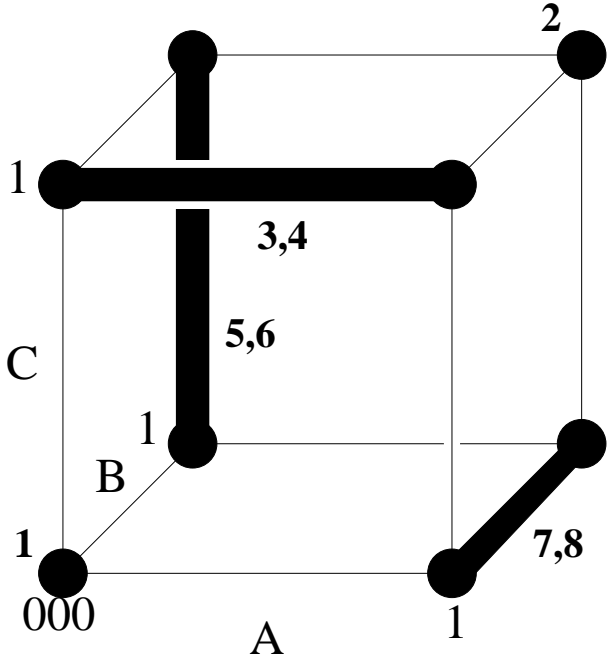}
\caption{The layout of the eight states of Eq. (\protect\ref{8states})
in the $2\times 2\times 2$ Hilbert space.  The ``dumbbells'' have a
similar meaning as the dominoes in Fig. \protect\ref{fig1}.}
\label{fig5}
\end{figure}

Note that nothing in the argument involves the simple product
states $\phi_1$ or $\phi_2$.  We conclude that the measurement is
still not doable locally even if these two states are promised to be
absent.  On the other hand, it is easy to show that eliminating any
one of the states $\phi_{3-8}$ would permit the measurement to be
done.  The layout of these states in the Hilbert space shown in
Fig. \ref{fig5} gives some intuition for why these should be true, as
in the two-party case: any simple von Neumann involves cutting one of
these ``dumbbells'' and making those pairs of states
indistinguishable.

Finally, the most economical technique that we have found for
making the measurement doable with quantum communication is for a
whole qubit to be sent from one party to another.  That is, no
compression of the quantum information seems to be possible in this
case, whether or not states $\phi_1$ or $\phi_2$ are excluded.  It is
easy to show that the resulting two-party measurement which is
required after this qubit transmission is doable by local actions.

\section{Discussion}
\label{disc}

The results of this paper, extensive as they are, raise many
additional fundamental questions about multipartite quantum
measurements, most of which we have only incomplete answers to at this
time.  We would indeed be pleased if the ambitious reader has a notion
of how to attack any of the following puzzles:

There are a variety of simple variants on the separable measurements
presented in this paper for which we do not know how to prove or
disprove bilocality.  One is a very obvious generalization of the
9-state example:
\begin{equation}
\begin{array}{lll}
\,&{\mbox \small Alice}&{\mbox \small Bob}\\
\psi'_1=&\;\;\;|1\rangle&\;\;\;|1\rangle,\\
\psi'_2=&\;\;\;|0\rangle&\;\;\;\cos\theta_{23}|0\rangle+
\sin\theta_{23}|1\rangle,\\
\psi'_3=&\;\;\;|0\rangle&-\sin\theta_{23}|0\rangle+
\cos\theta_{23}|1\rangle,\\
\psi'_4=&\;\;\;|2\rangle&\;\;\;\cos\theta_{45}|2\rangle+
\sin\theta_{45}|1\rangle,\\
\psi'_5=&\;\;\;|2\rangle&-\sin\theta_{45}|2\rangle+
\cos\theta_{45}|1\rangle,\\
\psi'_6=&\;\;\;\cos\theta_{67}|2\rangle+\sin\theta_{67}|1\rangle&
\;\;\;|0\rangle,\\
\psi'_7=&-\sin\theta_{67}|2\rangle+\cos\theta_{67}|1\rangle&
\;\;\;|0\rangle,\\
\psi'_8=&\;\;\;\cos\theta_{89}|0\rangle+\sin\theta_{89}|1\rangle&
\;\;\;|2\rangle,\\
\psi'_9=&-\sin\theta_{89}|0\rangle+\cos\theta_{89}|1\rangle&
\;\;\;|2\rangle.
\end{array}
\label{9pstates}
\end{equation}
That is, each of the domino pair is rotated by a different angle.
While we strongly doubt that the case of general $\theta$s is any
different from the case $\theta=\pi/4$ that we have analyzed, we have
no proof that these general states specify a nonlocal measurement.


We have noted that, although there is no $2\times 2$ pure-state
example that involves pure states of a separable but nonlocal
superoperator, there is a mixed-state measurement which has some very
curious properties.  It is a measurement to distinguish two density
matrices $\rho_0$ and $\rho_1$, where $\rho_0$ is an equal mixture of
the pure product states \verb$0+$ and \verb$+0$ (we use the notation
introduced in Eq. (\ref{8states})) and $\rho_1$ is an equal mixture of
\verb$11$ and \verb$--$.  It appears that, despite the fact that this
measurement involves distinguishing two separable, orthogonal states,
nevertheless the measurement cannot be done bilocally, indeed, the
measurement apparently cannot be done by any separable superoperator!
It is easy to show that the projection measurement into these states
can produce an entangled output from an unentangled input (for
instance, \verb$0$(\verb$0$+\verb$+$)); no separable superoperator can
do this.  It will be interesting to understand the minimum degree of
nonlocality needed to perform this measurement.

A nonlocal measurement would yield 1 bit of information since $\rho_0$
and $\rho_1$ are orthogonal.  It would be interesting to try to apply
the techniques developed in this paper to determine a bound on the
attainable mutual information by a bilocal approximation to this
measurement.

There are other multi-party examples for which such proofs would also
be desirable.  A modified version of the $2\times 2\times 2$ example
above involves just four states:
\begin{eqnarray}
&\verb%01+%&\nonumber\\
&\verb%1+0%&\nonumber\\
&\verb%+01%&\label{4states}\\
&\verb%---%.&\nonumber
\end{eqnarray}
These states do not correspond to a separable trace-preserving
superoperator, as the complement to these four measurements is not
separable\cite{foot4}.  Nevertheless, this can be viewed as a
measurement game in which Alice, Bob, and Carol are promised that they
are given one of these four states, and their object is to
distinguish, with only classical communications, which state it is.
We suspect that they cannot, but we have not been able to prove it.

An even more exotic set of orthogonal states that we have considered
is one involving 10 parties, each with a qubit.  This set of states
again only involves basis vectors \verb%0%, \verb%1%, \verb%+%, and
\verb%-% locally, so that a typical one of the 1024 basis states is
\verb%1+-+0--110%.  This construction emerges from a counterexample of
a proposition in tiling theory, the Keller conjecture\cite{Keller}.
The violation of this conjecture means that the 1024 states do not
conform to the domino or dumbbell layout of the examples in
this paper, where pairs of dimensions of the Hilbert space are covered
by pairs of orthogonal states.  We have not attempted to prove
non-10-locality for this example, but we note that there is no simple
von Neumann measurement that will distinguish them.

Curiously, despite the complexity of the example, we are able to show
that just two copies of any state are sufficient for the 10 parties to
be able to locally distinguish the state with classical communication,
as in all the examples considered in Sec. \ref{sec4}.  The procedure
is simple: measure one copy in the \verb%0/1% basis, and the second in
the \verb%+/-% basis.  This has raised another question: are there any
sets of states, entangled or not, for which some finite number greater
than 2 of copies of the state is necessary for distinguishing the
states reliably?  So far we have found no examples where more than two
copies of the unknown state are needed\cite{Pop}.  Indeed we know of
no examples of two orthogonal pure states, product or entangled, which
require more than {\em one\/} copy to be relibaly distinguished.
Earlier in this section we noted a set of two orthogonal {\em mixed\/}
states of two qubits which appears to be locally immeasurable. But
here too, two copies are sufficient to make the states
distinguishable.  It would appear that further work on the tiling
problems could produce other interesting examples for numbers of
parties between 3 and 10.

The domino representation of two-party quantum states bears some
resemblance to an approach taken in classical communication complexity
problems to finding the most efficient interactive scheme for
evaluating a function of data held by both Alice and Bob with the
minimum classical communication\cite{Hoy}.  The resemblance comes when
the one-bit output of the function is depicted in a two-dimensional
table; then the most efficient communication is determined by a
recursive subdivision of such a table into unanimous blocks.  It
remains to be seen whether this observation would lead to more
examples of interesting separable quantum operations.

The present investigation has required a very precise distinction
between different types of quantum operations which are normally
considered identical.  Going back to the 9-state calculation,
we can consider two different quantum operations related to the
measurement operation of Eq. ({\ref{supop}) (repeated here):
\begin{eqnarray}
&&|i\rangle_A|i\rangle_B\langle\psi_i|,\label{op1}\\
&&|\psi_i\psi_i\rangle\langle\psi_i|,\label{op2}\\
&&|\psi_i\rangle\langle\psi_i|.\label{op3}
\end{eqnarray}
We have disproved the existence of (\ref{op1}).  We can from this
disprove the existence of (\ref{op2}), which is a cloning operator: we
just note that Alice and Bob could perform this cloning many times,
then perform measurements to deduce with very high confidence the
state label $i$, thus performing (\ref{op1}).  We can also rule out
any form of weak cloning\cite{BBM}.  The case for (\ref{op3}) is more
subtle, since we normally think of these projection operators as
precisely what we mean by the measurement (\ref{op1}).  This is true
in a one-party world, since performing the projection means that a
classical record of the state is available somewhere in the world.
But in a multi-party situation, this record could be in a form which
is split between the parties in a way which would require quantum
communication to unravel.  Therefore, we emphatically state that
(\ref{op1}) and (\ref{op3}) are {\em not} generally identical in a
multi-party scenario.  Indeed, we note that there is another case in
which two such operators are completely different.  For Bell states,
the measurement operator Eq. (\ref{op1}) cannot be done bilocally,
because of the entanglement of the states; but the dephasing operator
Eq. (\ref{op3}) for the Bell states can be done bilocally; it has been
described as the ``twirling'' operation of Ref. \cite{BDSW}.

Nevertheless, we have been able to prove that (\ref{op3}) is not
doable for the 9-state examples, but by quite different arguments than
those given for Eq. (\ref{op1}), presented in Appendix \ref{appc}.
But the issue of approximations to (\ref{op3}), or (\ref{op2}),
remains unsettled.  That is, we do not know how to quantify the
precision with which Alice and Bob could do these operations
approximately.  A large part of the difficulty is that we cannot use a
simple, classical measure of information like the mutual information,
which was possible for (\ref{op1}) because the output is a classical
record.  For (\ref{op2}) and (\ref{op3}) an operator measure,
involving a notion of distance between two quantum operators, would
have to be used.  The theory of such operator measures is considerably
less well developed\cite{butsee}.

It seems likely that the states we have explored in this paper would
be usable for quantum cryptography, but we also have more questions
than answers on this point.  It is now clear\cite{Tal98} that
bipartite orthogonal states are generally useful for cryptography when
one particle in the state is received by Bob before the other has been
launched by Alice.  This forces Eve to measure one particle at a time.
If Eve had no quantum memory, then the security of the cryptography
protocol would be assured if the measurement of the state could not be
performed bilocally, with the restriction that only round of
measurement (one transmission from Alice to Bob) would be permitted.
The nine states that we have analyzed have this property.  However,
given that Eve can have a quantum memory, the problem is a bit
different, corresponding to there being some restricted form of
quantum communication between Alice and Bob in the measurement
protocol.  In the cryptographic application, of course, Eve has more
work to do: she must determine the identity of the state {\em and}
provide it undisturbed, at the appropriate times, to
Bob\cite{Fuchs96a}.  Thus, a separate study is required to establish
that the nine states form a good basis for orthogonal quantum
cryptography (which, however, is easily provided by the analysis of
\cite{Tal98}).  At the same time, we may imagine that the nine states
might provide a stronger cryptographic primitive for some purposes,
given that they cannot be identified even by repeated communication
between Alice and Bob.  Another useful feature of the states as a
cryptographic primitive might be the fact that two copies of them can
be identified exactly.  But we have no concrete notions of what these
new cryptographic applications might be.

Finally, we note that the basic question which began our investigation
remains unanswered: what is a compact mathematical description of
a superoperator which can be performed by only classical communication
between the parties?  We have only disproved one natural hypothesis,
that this set coincides with the set of separable superoperators.
No alternative hypothesis has presented itself.

All of these questions indicate, we think, that we still have many
very basic questions about the structure of quantum mechanics, about
the nature of quantum nonlocality and of entanglement, questions
whose answers will be of central significance in our quest to employ
quantum mechanics in the transmission and processing of information.

\section*{acknowledgments}

Part of this work was completed during the 1997 Elsag-Bailey---I.S.I.
Foundation research meeting on quantum computation.  CAF has been
supported by a Lee A. DuBridge Fellowship and by DARPA through the
Quantum Information and Computing (QUIC) Institute administered by the
US Army Research Office.  We thank Micha{\l} Horodecki, Peter H{\o}yer,
N. David Mermin, Sandu Popescu, Barbara Terhal, and Reinhard Werner
for very helpful discussions.

\appendix

\section{Decomposition of arbitrary POVM into a series of very weak
measurements}
\label{peter}

Any superoperator acting on a system of dimension $n$ can be replaced
(nonuniquely) by the following procedure: appending an ancilla of
dimension $n_1$, performing a unitary transformation, tracing out a
subsystem of dimension $n_2$, and measuring (using a standard and
complete measurement) a subsystem of dimension $n_3$, which we call a
probe.  As a result, the state of remaining system (of dimension $m =
\frac{n n_1}{n_2 n_3}$) can be calculated, and it is uniquely
determined for any given superoperator despite of the nonuniqueness of
the procedure.  In cases where there is no probe to be measured
($n_2=1$), this is the so-called trace-preserving superoperator.  If
instead the trace-out step is eliminated, this is the most general
POVM (positive operator-valued measurement).  In our case, where all
information is used for the optimal extraction of information, we are
interested in this second case.  Thus, the most general POVM can be
replaced by the three operations -- the appending of an ancilla, the
unitary transformation and the standard measurement of a subsystem.

Suppose we are given a state, on which we will obtain some information
using a POVM.  We will show how to approximate this POVM by a
continuous process.  The addition of ancilla does not influence the
state; the unitary transformation can be done as continuously as we
wish.  We shall now show that a standard complete measurement can be
replaced by a continuous process (to any desired approximation).  As a
result of the above discussion, any POVM can be approximated in the
same way.

In order to measure the probe (a subsystem of dimension $n_3$) using a
complete standard measurement in a basis $|i\rangle$, we write the
combined state of the remaining system and the probe (of dimensions
$nn_1 = mn_3$) after the unitary interaction as
\begin{equation}
|\psi\rangle = \sum_{i=0}^{n_3-1} \alpha_i | \phi_i \rangle |i\rangle \,
\end{equation}
where $\sum_{i=0}^{n_3-1} |\alpha_i|^2 = 1$, and where
$|\phi_i\rangle$ are normalized states (not necessarily orthogonal)
of the remaining $m$-dimensional subsystem.  Without loss of
generality we can assume that the probe is a qubit, since any other
measurement can be replaced by a set of yes/no questions, thus
$n_3=2$.

In a standard measurement we apply the projection postulate directly
on the probe to yield a classical result $i$ with probability
$|\alpha_i|^2$, and a remaining subsystem in a state $|\phi_i\rangle$.
In a nondemolition measurement\cite{Caves} a state $|i\rangle$ is
tranformed to $|i\rangle |i\rangle_1$ and the new system
($|i\rangle_1$) is measured instead of the probe, hence a probe in a
state $|i\rangle$ is not demolished by this measurement.  Attaching
$K$ such devices to $|\psi\rangle$, the measurement of the probe can
be done in a nondemolition way using a unitary transformation to a
state
\begin{equation}
| \Phi\rangle = \sum_{i=0}^1 \alpha_i | \phi_i \rangle |i\rangle \
|i\rangle_1 |i\rangle_2 \ldots |i\rangle_K \ ,\label{ndstate}
\end{equation}
where now the measurement postulate can be applied on any (or on all)
of the additional ``quantum measuring devices'' $|i\rangle_k$, where
$1\le k \le K$.  We use the term quantum measuring device (QMD) to say
that no classical measurement (no actual ``printout'') was performed
at that stage.  As a result, this measurement process is reversible
until we apply the projection postulate on one of these QMDs, and the
state $|\psi\rangle$ can be reproduced from $|\Phi\rangle$ with
perfect fidelity.  Measuring any of these QMDs is equivalent to
performing a standard measurement on $|i\rangle$.

To obtain an approximation using a continuous measurement we replace
the QMDs by ``weak QMDs'' (WQMDs), meaning that we replace a standard
measurement by a sequence of weak measurements.
(Weak measurements were first discussed by Aharonov and
others~\cite{Aharonov}.)
The unitary transformation producing (\ref{ndstate}) is replaced by
one leading to
\begin{equation}
| \Psi\rangle =
\alpha_0 | \phi_0 \rangle |0\rangle \ |0'\rangle_1
\ldots |0'\rangle_K +
\alpha_1 | \phi_1 \rangle |1\rangle \ |1'\rangle_1
\ldots |1'\rangle_K \ ,
\end{equation}
where the two possible states of the $k$'th WQMD, $|0'\rangle_k$ and
$|1'\rangle_k$ are highly overlapping.  We can always choose them to
be
\begin{eqnarray}
&&|0'\rangle = \cos \theta |0\rangle + \sin \theta |1\rangle,\nonumber\\
&&|1'\rangle = \sin \theta |0\rangle + \cos \theta |1\rangle,
\end{eqnarray}
with $\theta = \pi/4 - \epsilon$ with small positive $\epsilon$.  If
the state we start with is $|i\rangle$, then the probability to obtain
a correct result $i$ from a probe in a state $|i'\rangle$ is
\begin{equation}
\cos^2 \theta = 1/2 [1 + \sin (2\epsilon)].
\end{equation}
We approximate:
\begin{eqnarray}
&&\cos \theta \approx (1/\sqrt 2) [1 + \sin (2\epsilon)/2] \approx
(1/\sqrt 2) [1 + \epsilon],\nonumber\\
&&\sin \theta \approx (1/\sqrt 2)[1-\epsilon].
\end{eqnarray}

For any state $|\psi\rangle$, if only one WQMD is measured (in the
computation basis), the effect of this measurement on the rest of the
system is weak, and the state of the original system can be reproduced
with high fidelity which approaches one as $\epsilon$ approaches zero.
For instance, if a result $0$ is obtained, we can reproduce an
unnormalized state of the remaining system and the probe
\begin{equation}
|\psi_{out}\rangle = \alpha_0 \cos \theta | \phi_0 \rangle |0\rangle
+ \alpha_1 \sin \theta | \phi_1 \rangle |1\rangle
\end{equation}
yielding a modification of $|\psi\rangle$ of order $\epsilon$:
\begin{equation}
| \psi_{out}\rangle = |\psi\rangle + \epsilon [\alpha_0 |\phi_0
\rangle |0\rangle - \alpha_1 |\phi_1 \rangle |1\rangle ] \ .
\end{equation}
Thus, measuring each such QMDs one at a time, we obtain a process
which is as close to continuous as we want, since we can choose
$\epsilon$ as small as we want.

The last thing to verify is that we can choose $K$ big enough
in order to yield the same probability of obtaining the result
$i$ as in a standard measurement.

If the state of the probe is $|i\rangle$, then each of the WQMDs is in
pure state $|i'\rangle$.  When we measure $K$ WQMDs their outcomes are
independent and identically distributed according to a binomial
distribution with probability $\cos^2\theta$ to obtain the correct
result $i$ for each one.  Let us assume that $K$ is odd.  When we look
at $K$ such WQMDs and take a majority vote, the probability to obtain
a correct result is given by
\begin{equation}
\sum_{k=1}^{(K-1)/2} {K \choose k}\cos^{2(K-k)} \theta \sin^{2k}
\theta.
\end{equation}
(Note that this expression
can also be calculated by expanding
\begin{equation}
|\phi_i\rangle |i\rangle \ |i'\rangle_1 \ldots |i'\rangle_K
=|\phi_i\rangle |i\rangle \ [ \cos^{K} \theta | i \ldots i \rangle +
\cos^{K-1} \theta \sin \theta | ii \ldots ij \rangle + \sin^{K} \theta
|jj \ldots j \rangle]$, with $j = 0
\end{equation}
if $i = 1$ and vice versa, and calculating the probability of each
string.)

This is equivalent to a classical problem of having a biased coin with
a known bias $\cos^2 \theta$, and trying to guess whether it is biased
to give more heads or more tails.  One can bound the above sum
directly, or approximate it using some central limit theorem (since it
is a random walk).

Alternatively, one can use a strong version of the law of large
numbers, which tells us that we can guess the direction of the bias
with probability exponentially close to one.  Suppose one throws a
biased coin, so that in one try it gives $Prob(x=1) = p$ and
$Prob(x=0) = 1-p$.  According to Bernstein law of large
numbers~\cite{Ber}, when throwing the same coin $K$ times the actual
average of the $K$ trials, $\sum_{i=1}^K x_i/K$, is very close to the
expectation value $p$, except with probability
\begin{equation}
Prob[| \sum_{i=1}^K x_i/K - p | \ge \delta] \le 2 e^{-K \delta^2}
\end{equation}
for any $K$, and for $\delta$ smaller than $p(1-p)$.

To apply this law to our case recall $\cos^2\theta = 1/2 + \sin
2\epsilon / 2$, and $\sin^2\theta = 1/2 - \sin 2\epsilon / 2$, so that
Bernstein law applies for any $\delta < 1/4 - \sin^2 2\epsilon$.  For
small $\epsilon$ (e.g. less than 1/8) we choose $\delta = \sin 2
\epsilon /2$ which is in the appropriate range.  Now, the probability
of observing $\sum_{i=1}^K x_i/K \ge 1/2 $ when the $Prob (x=1) =
\sin^2\theta$ is less than or equal to
\begin{equation}
Prob[| \sum_{i=1}^K x_i/K - p | \ge \sin 2\epsilon/2
] \le 2 e^{-K \sin^2 2\epsilon/4}.
\end{equation}
Since $K$ can be chosen independently of $\epsilon$, any $K \gg
4/\sin^2 2\epsilon$ will do.

This means that for such $K$ the expression
\begin{equation}
S = \sum_{k=1}^{(K-1)/2} {K \choose k} \cos^{2(K-k)} \theta \sin^{2k}
\theta
\end{equation}
is exponentially close to 1, and its complement
\begin{equation}
1-S = \sum_{k=1}^{(K-1)/2} {K \choose k} \sin^{2(K-k)} \theta
\cos^{2k} \theta
\end{equation}
is exponentially small.

In the general case of a state $|\Psi\rangle$, we need to expand the
state (as was done above), and calculate the probability of each
string in order to take a majority vote as before.  This process
yields (assuming as before odd $K$) a probability of
\begin{equation}
|\alpha_i|^2 \sum_{k=1}^{(K-1)/2} {K \choose k}
\cos^{2(K-k)} \theta  \sin^{2k} \theta
+ (1-|\alpha_j|^2) \sum_{k=1}^{(K-1)/2} {K \choose k}
\sin^{2(K-k)} \theta  \cos^{2k} \theta
\end{equation}
to obtain the correct result.
Using $S$ we get
\begin{equation}
|\alpha_i|^2 S + (1-|\alpha_j|^2) (1-S)
\end{equation}
so the result is obtained with the correct probability $ |\alpha_i|^2
S $.  (This is equivalent to obtaining a coin with bias $\cos^2\theta$
with probability $|\alpha_0|^2$ or with another bias $\sin^2\theta$
with probability $1-|\alpha|^2$ and throwing it as many times as we
want in order to learn which coin we received with any desired
probability of success.)

\section{Constraints from approximate orthogonality of residual states}
\label{appaa}

According to Eq. (\ref{orcon}) the overlaps between the residual
states $\phi_{i,m_I}$ after stage I (Eq. (\ref{residuary})) are all
bounded by $\delta$:
\begin{equation}
|\langle\phi_i|\phi_j\rangle|= {{|\langle\psi_i|a\otimes
b|\psi_j\rangle|}\over {\sqrt{\langle\psi_i|a\otimes b|\psi_i\rangle
\langle\psi_j|a\otimes b|\psi_j\rangle}}}\leq\delta,\;\;\;\; \forall
i\neq j.
\label{orcona}
\end{equation}
The task here is to use these inequalities to derive various
constraints on the matrix elements of the operators $a$ and $b$ in
Eq.~(\ref{aandb}).

We note before we begin that during the completion of stage I, Alice
and Bob may each have augmented their Hilbert spaces beyond their
original three dimensions.  They might do this, for instance, as part
of a strategy that requires retaining some of the quantum ancillae
from one round of the protocol to the next.  Such a strategy finds its
expression in the fact that the $S_{m_I}$ operators need not be square
matrices, so that the states of Eq. (\ref{residuary}) will live in a
Hilbert space larger than the original 9-dimensional one.
Fortunately, this contingency has no relevance for the constraints we
are about to derive: it is only the algebraic properties of
$E=S^\dagger S=a\otimes b$ that concern us, and $a$ and $b$ are always
square matrices whose dimensions are determined by the size of the
{\em initial\/} Hilbert space.

Let us use the notation $\langle i|a|j\rangle=a_{ij}$ and $\langle
i|b|j\rangle=b_{ij}$ and note the following preliminary things.
Recall that $a$ and $b$ are both positive semidefinite operators so
that, for each $i$, $\langle\psi_i|a\otimes b|\psi_i\rangle>0$.
Recall that, from Eq.~(\ref{bound}), we have for each $i$ and $j$,
\begin{equation}
0<\frac{\langle\psi_i|a\otimes b|\psi_i\rangle}%
{\langle\psi_j|a\otimes b|\psi_j\rangle}\le\frac{1+9\epsilon}%
{1-72\epsilon}\;.
\label{Ecclesiastes}
\end{equation}
Thus it follows that $a_{00}$, $a_{22}$, $b_{00}$, and $b_{22}$ are
all strictly positive.  Moreover, looking at $i=2$ and $j=3$, for
instance, in Eq.~(\ref{Ecclesiastes}) gives
\begin{equation}
0<{{b_{00}+b_{11}+2{\rm Re}\,b_{10}}\over
{b_{00}+b_{11}-2{\rm Re}\,b_{10}}}\le\frac{1+9\epsilon}
{1-72\epsilon}\;.
\end{equation}
From this and the $i=3$, $j=2$ condition:
\begin{equation}
|2{\rm Re}\,b_{10}|\le\frac{81\epsilon}{2-63\epsilon}(b_{11}+b_{00})
\;.
\label{Smurfa}
\end{equation}
In similar fashion, taking $\{i,j\}=\{2k,2k+1\}$ for $k=2,3,4$, we
have
\begin{eqnarray}
|2{\rm Re}\,b_{21}|
&\le&
\frac{81\epsilon}{2-63\epsilon}(b_{22}+b_{11})\;,
\label{Murfa}
\\
|2{\rm Re}\,a_{21}|
&\le&
\frac{81\epsilon}{2-63\epsilon}(a_{22}+a_{11})\;,
\\
|2{\rm Re}\,a_{10}|
&\le&
\frac{81\epsilon}{2-63\epsilon}(a_{11}+a_{00})\;.
\end{eqnarray}

We can now bound the relative variations among the diagonal elements
of $a$ and $b$ in terms of $\delta$ in the following way.  Taking
$i=2$ and $j=3$, Eq.~(\ref{orcona}) looks like
\begin{equation}
{{a_{00}|b_{00}-b_{01}+b_{10}-b_{11}|}\over
{\sqrt{a_{00}(b_{00}+b_{01}+b_{10}+b_{11})
a_{00}(b_{00}-b_{01}-b_{10}+b_{11})}}}\leq\delta\;,
\label{bineq}
\end{equation}
and simplifies to
\begin{equation}
{{|b_{00}-b_{11}+2i{\rm Im}\,b_{10}|}\over{\sqrt{(b_{00}+b_{11})^2-
(2{\rm Re}\,b_{10})^2}}}\leq\delta\;.
\label{int2}
\end{equation}
Therefore it follows that
\begin{equation}
{{|b_{00}-b_{11}|}\over{b_{00}+b_{11}}}\leq\delta\;.
\end{equation}
In similar fashion, we get
\begin{equation}
{{|b_{11}-b_{22}|}\over{b_{11}+b_{22}}}\leq\delta\;,\qquad
{{|a_{11}-a_{22}|}\over{a_{11}+a_{22}}}\leq\delta\;,\qquad
{{|a_{00}-a_{11}|}\over{a_{00}+a_{11}}}\leq\delta\;.
\label{AsiaMinor}
\end{equation}

These inequalities help us bound the off-diagonal matrix elements of
$a$ and $b$ in terms of $\delta$.  Consider the combination of the
conditions from Eq.~(\ref{orcona}) given by taking $i=2,3$ and
$j=4,5$.  The $i=2,\,j=4$ inequality, for instance, is
\begin{equation}
{{|a_{02}(b_{02}+b_{01}+b_{12}+b_{11})|}\over{\sqrt{a_{00}a_{22}
(b_{00}+b_{11}+2{\rm Re}\,b_{10})(b_{22}+b_{11}+2{\rm Re}\,b_{21})}}}
\leq\delta\;.
\label{abse}
\end{equation}
It will be convenient to introduce the notation
\begin{equation}
D^2_{\pm\pm}=a_{00}a_{22}\Big((b_{00}+b_{11})\pm
2{\rm Re}\,b_{10}\Big)
\Big((b_{11}+b_{22})\pm2{\rm Re}\,b_{21}\Big)\;.
\end{equation}
With this, we see that we can write
\begin{equation}
a_{02}(b_{02}+b_{01}+b_{12}+b_{11})=
D_{++}\,\gamma^{(1)} e^{i\phi^{(1)}}\;,
\label{Muenchen}
\end{equation}
where $\gamma^{(1)}$ is a small amplitude and $\phi^{(1)}$ is an
appropriately chosen phase that satisfy the constraints
\begin{equation}
0\leq\gamma^{(1)}\leq\delta\quad\mbox{and}\quad 0\leq\phi^{(1)}<2\pi\;.
\label{Zorneding}
\end{equation}
In similar fashion, taking the remaining combinations of $i=2,3$ and
$j=4,5$ we arrive at
\begin{eqnarray}
a_{02}(-b_{02}+b_{01}-b_{12}+b_{11})
&=&
D_{+-}\,\gamma^{(2)} e^{i\phi^{(2)}}\;,
\label{Schwabish Hall}
\\
a_{02}(-b_{02}-b_{01}+b_{12}+b_{11})
&=&
D_{-+}\,\gamma^{(3)} e^{i\phi^{(3)}}\;,
\\
a_{02}(b_{02}-b_{01}-b_{12}+b_{11})
&=&
D_{--}\,\gamma^{(4)} e^{i\phi^{(4)}}\;,
\label{Lowenbraukeller}
\end{eqnarray}
where each $\gamma^{(i)}$ and $\phi^{(i)}$ satisfy the same
constraints as in Eq.~(\ref{Zorneding}).  Adding
Eqs.~(\ref{Muenchen}) and
(\ref{Schwabish Hall})--(\ref{Lowenbraukeller}) together and taking
the absolute value of the resultant, we get
\begin{equation}
4|a_{02}b_{11}|\le\delta(D_{++}+D_{+-}+D_{-+}+D_{--})
\label{Dilbert}
\end{equation}
Now suppose that ${\rm Re}\,b_{10}\ge0$ and ${\rm Re}\,b_{21}\ge0$,
and let
\begin{equation}
z=1+\frac{81\epsilon}{2-63\epsilon}=
\frac{2+18\epsilon}{2-63\epsilon}\;.
\end{equation}
Then it follows from Eqs.~(\ref{Smurfa}) and (\ref{Murfa}) that
\begin{eqnarray}
D^2_{++}
&\le&
z^2 a_{00}a_{22}(b_{00}+b_{11})(b_{11}+b_{22})\;,
\\
D^2_{+-}
&\le&
z a_{00}a_{22}(b_{00}+b_{11})(b_{11}+b_{22})\;,
\\
D^2_{-+}
&\le&
z a_{00}a_{22}(b_{00}+b_{11})(b_{11}+b_{22})\;,
\\
D^2_{--}
&\le&
a_{00}a_{22}(b_{00}+b_{11})(b_{11}+b_{22})\;.
\end{eqnarray}
Combining this with Eq.~(\ref{Dilbert}), we find
\begin{equation}
\frac{|a_{02}|}{\sqrt{a_{00}a_{22}}}\le\frac{1}{4}\delta
\left(z+2\sqrt{z}+1\right)\frac{1}{b_{11}}
\sqrt{(b_{00}+b_{11})(b_{11}+b_{22})}\;.
\label{Hurrumph}
\end{equation}
Note that Eq.~(\ref{Hurrumph}) remains true regardless of
the signs of ${\rm Re}\,b_{10}$ and ${\rm Re}\,b_{21}$.  This is
because Eq.~(\ref{Dilbert}) remains invariant under a change of sign
for either or both of these terms.  Now it is just a question of
using the previously derived constraints for the diagonal elements of
$a$ and $b$ to put a limit on how large the right-hand side of this
can be.  With some play, one sees that this occurs when
\begin{equation}
b_{00}=b_{22}=b_{11}\frac{1+\delta}{1-\delta}\;,
\end{equation}
and, at that point, one has
\begin{equation}
\frac{|a_{02}|}{\sqrt{a_{00}a_{22}}}\le\frac{1}{2}
\left(z+2\sqrt{z}+1\right)\frac{\delta}{1-\delta}\;.
\label{BuckNaked}
\end{equation}
Alternatively taking $i=6,7$ and $j=8,9$ in Eq.~(\ref{orcona}) and
running through a set of steps analogous to those in
Eqs.~(\ref{abse}) through (\ref{BuckNaked}), one finds
\begin{equation}
\frac{|b_{02}|}{\sqrt{b_{00}b_{22}}}\le\frac{1}{2}
\left(z+2\sqrt{z}+1\right)\frac{\delta}{1-\delta}\;.
\end{equation}

By a slightly more elaborate strategy, we can now find bounds on
all the remaining off-diagonal terms.  Let us consider the
inequalities derived from Eq.~(\ref{orcona}) for $i=4,5$ and $j=8,9$.
These can all be written in a compact notation as
\begin{equation}
{{|(a_{20}+(-1)^{s_1}a_{21})(b_{22}+(-1)^{s_2}b_{12})|}\over
{\sqrt{a_{22}
(a_{00}+a_{11}+(-1)^{s_1}2{\rm Re}\,a_{01})
(b_{22}+b_{11}+(-1)^{s_2}2{\rm Re}\,b_{12})
b_{22}}}}\leq\delta\;.
\label{numfac}
\end{equation}
The sign bit $s_1=0,1$ corresponds to $j=8,9$; the bit $s_2=0,1$
corresponds to $i=4,5$. Let us focus on only one of these four
equations, one for which
\begin{eqnarray}
(-1)^{s_1}{\rm Re}\,a_{01} &\leq& 0\;,
\label{c1}
\\
(-1)^{s_2}{\rm Re}\,b_{12} &\geq& 0\;.
\label{c2}
\end{eqnarray}
It is clear that at least one of the four sign choices will satisfy
these conditions.  In that case, it follows that
\begin{equation}
{{|a_{20}+(-1)^{s_1}a_{21}|}\over{\sqrt{a_{22}
(a_{00}+a_{11})}}}\leq\delta\sqrt{\frac{1}{b_{22}}\left(b_{22}+b_{11}+
(-1)^{s_2}2{\rm Re}\,b_{12}\right)}\;.
\end{equation}
Using Eq.~(\ref{Murfa}), this implies
\begin{equation}
{{|a_{20}+(-1)^{s_1}a_{21}|}\over{\sqrt{a_{22}
(a_{00}+a_{11})}}}\leq\sqrt{z}\delta
\sqrt{\frac{1}{b_{22}}(b_{22}+b_{11})}\;.
\end{equation}
Maximizing the right-hand side of this subject to the constraint
Eq.~(\ref{AsiaMinor}) gives that
\begin{equation}
{{a_{20}+(-1)^{s_1}a_{21}}\over\sqrt{a_{22}(a_{00}+a_{11})}}=
\nu_1 e^{i\phi^{(5)}},\qquad 0\leq\nu_1\leq\delta\sqrt{{2z}\over
{1-\delta}}\;.
\label{gooda}
\end{equation}
Hence, using Eq.~(\ref{BuckNaked})
\begin{eqnarray}
{{|a_{21}|}\over{\sqrt{a_{22}(a_{00}+a_{11})}}}
&\le&
\delta\sqrt{{2z}\over{1-\delta}}+{{|a_{20}|}\over{\sqrt{a_{22}
(a_{00}+a_{11})}}}
\\
&\le&
\delta\sqrt{{2z}\over{1-\delta}}+\frac{1}{2}
\left(z+2\sqrt{z}+1\right)\frac{\delta}{1-\delta}
\sqrt{\frac{a_{00}}{a_{00}+a_{11}}}\;.
\end{eqnarray}
Finally optimizing the left and right-hand sides of this subject to
the constraints imposed on $a_{00}$ by Eq.~(\ref{AsiaMinor}),
we obtain
\begin{equation}
{{a_{21}}\over{\sqrt{a_{22}a_{11}}}}=\nu_2
e^{i\phi^{(6)}},\qquad 0\leq\nu_2\leq\nu_{\epsilon}=
{{2\delta}\over{1-\delta}}\left(\sqrt{z}+\frac{1}{4}
\left(z+2\sqrt{z}+1\right)\sqrt{\frac{1+\delta}{1-\delta}}\right).
\label{thebiggie}
\end{equation}
This is the desired bound.  Applying exactly the same reasoning to
Eq.~(\ref{orcona}) with $i=6,7$ and $j=4,5$, we find the same bound
on $|b_{10}|/\sqrt{b_{11}b_{00}}$.  Doing the same with $i=2,3$ and
$j=6,7$, we find it for $|a_{01}|/\sqrt{a_{00}b_{11}}$; and finally,
with $i=8,9$ and $j=2,3$, we find it for
$|b_{12}|/\sqrt{b_{11}b_{22}}$.

\section{Compressibility of classical advice}
\label{appd}

To see how negative advice of the form ``not state $j$'' can be
asymptotically compressed, consider first the simple case of the
equiprobable 8-state ensemble.  Suppose Alice and Bob are faced with
the task of performing a large number $n$ of the 8-state measurements;
they are promised that $\psi_1$ does not occur and all other states
are equi-probable (this is the simplest case).  Then they must
ultimately distinguish $8^n$ possible outcomes.  But one single string
of $n$ hints (e.g., state 1 is not $\psi_2$, state 2 is not $\psi_9$,
state 3 is not $\psi_2$, etc.)  successfully covers $7^n$ of the
possible outcomes.  Thus, only approximately $({8\over 7})^n$ distinct
hint strings need ever be used to help Alice and Bob with their
measurements.  If Alice, Bob and the hint-giver pre-agree on which
hint strings are to be used and agree on a numbering of them (which
amounts to the selection of an expanding hash function), then the hint
can be conveyed in $n\log_2{8\over 7}$ bits, or $\log_2{8\over
7}\approx 0.193$ bits per measurement.

For general, not necessarily equal, prior probabilities $p_i$ of the 9
states, more sophisticated counting methods are required to calculate
compressibility of the hints. Let $X$ be a typical sequence of $n$
states chosen independently with probabilities $\{p_i, i=1...9\}$,
having about $np_i$ states of type $i$ for each $i$.  Let $Y$ be a
sequence of $n$ hints of the form ``the state is not state $j$'' is
chosen independently with probabilities $\{q_j, j=2...9\}$. A hint
sequence $Y$ is valid for a state sequence $X$ if none of the hints is
false (e.g., if $X=136$, then $Y=353$ is valid but $Y=356$ is not,
because the last hint is false). The probability that the hint
sequence $Y$ will be a valid for state sequence $X$ is
$\Pi_{i=2}^9p_i^{n(1-q_i)}$; therefore, using an expanding hash
function from an appropriate strongly 2-universal class\cite{Weg}, one
can show that $-\sum_{i=2}^9 p_i\log_2(1-q_i)$ bits of advice per
state are asymptotically necessary and sufficient to specify a valid
hint sequence for a typical X. The optimal compression for hints of
this sort can then be obtained by varying the probabilities $q_i$ to
minimize the above expression. When this is done, it turns out that if
one or more of the states $p_i$ occurs with probability significantly
higher than average, the corresponding hint ``not $p_i$'' should never
be used, i.e., the corresponding hint probability $q_i$ vanishes.

\section{Dephasing superoperator ruled out by invertibility result}
\label{appc}

In this Appendix we show that the superoperator which dephases in the
nine-state basis, Eq. (\ref{op3}), cannot be implemented bilocally by
Alice and Bob.  As a preliminary, we prove that this superoperator
cannot be performed with no classical communication between Alice and
Bob.  Consider two possible input states to the superoperator,
${1\over\sqrt{2}}(\psi_8-\psi_9)$ and $\psi_1$ (see Eq.
(\ref{9states})).  These states have the same reduced density operator
for Alice, so if there is no communication the output states must have
the same reduced density operator; but the dephasing superoperator
requires that they be different (in the first case ${1\over
2}(\ket{0}\bra{0}+\ket{1}\bra{1})$, and in the second
$\ket{1}\bra{1}$).  Thus, the superoperator is not doable without
communication.

Now we consider the case where some data stream $m$ (see Eq.
(\ref{concat})) has passed between Alice and Bob.  Without loss of
generality, we can assume that all the data transmission occurs before
Alice and Bob trace out any of the ancilla Hilbert spaces that they
have introduced (recall that the output space of Eq. (\ref{op3}) must
be the same $3\times 3$ space as the input).  Now, adopting the
``continuumized" view of superoperators that involve channel
transmissions introduced in Sec. \ref{continn}, we proceed with the
proof by considering two separate cases: 1) If the input to the
superoperator is one of the nine states $\psi_i$, the set of residual
states $S_m\ket{\psi_i}$ (Eq. (\ref{residuary})) at a certain instant
become nonorthogonal, without any of the states being annihilated
(non-annihilation is an obvious requirement of the dephasing
superoperator).  2) The residual states always remain orthogonal.

1) Disproving the bilocality of the dephasing superoperator in the
case where residual states become nonorthogonal is accomplished by
the following discussion of {\em invertibility} for superoperators.

Definition: A superoperator ${\cal S} = \{S_i\}$ is {\it weakly
invertible} relative to a set of pure states $\{\ket{v_k}\}$ if there
exist superoperators ${\cal T}_i = \{T_{ij}\}$ for each $i$ such that
the superoperator ${\cal U} = \{T_{ij} S_i\}$ satisfies
\begin{equation}
{\cal U}(\ket{v_k}\bra{v_k}) = \ket{v_k}\bra{v_k}\label{theinverse}
\end{equation}
for all $k$.  Note that the conventional projection superoperator of
Eq. (\ref{op3}) is one such operator of the form $\cal U$.

Since the dephasing operator Eq. (\ref{op3}) is an example of an
operator of the form of ${\cal U}$ in Eq. (\ref{theinverse}), any
partial completion of the superoperator up to some instant, in
particular the instant at which the residual states become
nonorthogonal, must be weakly invertible.  But we can easily
contradict this with the following lemma:

{\em If the superoperator ${\cal S}=\{S_i\}$ is weakly invertible
relative to the set $V$ of pure states, then for all $\ket{v_1}$,
$\ket{v_2} \in V$, if $\ket{v_1}$ and $\ket{v_2}$ are orthogonal, then
so are $S_i\ket{v_1}$ and $S_i\ket{v_2}$ for all $i$.}

Proof: Let ${\cal T}_i$ be superoperators demonstrating the weak
invertibility of ${\cal S}$.  Then, by definition, for all $\ket{v}$
in $V$,
\begin{equation}
\sum_{ij} T_{ij} S_i \ket{v}\bra{v} S^\dagger_i T^\dagger_{ij}
=\ket{v}\bra{v}.
\end{equation}
This implies that for all $i$, $j$,
\begin{equation}
T_{ij} S_i\ket{v} = \alpha_{ij}(v) \ket{v},
\end{equation}
for some scalar $\alpha_{ij}(v)$.  But then
\begin{equation}
(T_{ij} S_i\ket{v_2})^\dagger (T_{ij} S_i \ket{v_1}) =
\alpha_{ij}(v_1) (\alpha_{ij}(v_2))^* \langle v_2|v_1\rangle =
0,
\end{equation}
so
\begin{equation}
\sum_j (T_{ij} S_i \ket{v_2})^\dagger (T_{ij} S_i \ket{v_1})
= \bra{v_2}S_i^\dagger S_i\ket{v_1} = 0.
\end{equation}

2) Disproving the bilocality of the dephasing superoperator in the
case where residual states always remain orthogonal throughout the
period when Alice and Bob are communicating through the channel
requires a different line of argument from case 1.  First, we note
that the calculation of Appendix \ref{appaa} shows that if the states
remain exactly orthogonal (cf. Eq. (\ref{orcona})), then each operator
$a_m$ and $b_m$ must be exactly proportional to the identity operator;
this in turn implies that each operation element is proportional to a
product of an Alice and a Bob unitary operator,
\begin{equation}
S_m=\alpha_mU_{mA}\otimes U_{mB}.\label{prods}
\end{equation}
Note that by the conditions of Appendix \ref{appaa}, the posterior
probabilities must remain finite for this result to hold; but, as
noted before, if this condition were not satisfied, it could be
immediately argued that the superoperator could never result in the
desired dephasing operator.  In fact, of course, using
Eqs. (\ref{bys},\ref{stan}) it is straightforward to show that the
posterior probabilities $p(\psi_i|m)$ remain identical to the prior
probabilities $p(\psi_i)$; no information about the states ever flows
through the classical channel.

Given that the superoperator is constrained to be of the form
Eq. (\ref{prods}), it is easy to complete the proof.  (\ref{prods})
implies, for each state $\psi_i$ of Eq. (\ref{9states}),
\begin{equation}
U_{mA}\otimes U_{mB}|\alpha_i\otimes\beta_i\rangle=e^{i\theta_i}
|\alpha_i\otimes\beta_i\rangle.
\end{equation}
This implies
\begin{equation}
U_{mA}|\alpha_i\rangle=e^{i\theta_{iA}}|\alpha_i\rangle,\label{phaq}
\end{equation}
and a similar relation for $B$.  (It is this last step which cannot be
taken for the Bell-state dephasing case mentioned in the text.)  Now,
referring to Eq. (\ref{9states}), considering cases $i=1,2,4$ shows
that $U_{mA}$ is diagonal in the $|0,1,2\rangle$ basis; then $i=8,9$
shows that $\langle0|U_{mA}|0\rangle=\langle1|U_{mA}|1\rangle$, and
$i=6,7$ shows that
$\langle1|U_{mA}|1\rangle=\langle2|U_{mA}|2\rangle$.  Thus, $U_{mA}$,
and similarly $U_{mB}$, are proportional to the identity operator.
But the identity superoperator can be done without any classical
communication, and the argument at the beginning of this appendix
shows how this possibility is excluded.

This completes the proof for both cases.


\begin{references}

\bibitem{ekert} A. K. Ekert, ``Quantum cryptography based on Bell's
theorem,'' Phys. Rev. Lett. {\bf 67}, 661 (1991).

\bibitem{bfs} C.H. Bennett, C.A. Fuchs, and J.A. Smolin,
``Entanglement-Enhanced Classical Communication on a Noisy Quantum
Channel", in {\em Quantum Communication, Computing, and
Measurement} (eds. O. Hirota, A.S. Holevo, and C.M. Caves, Plenum, New
York, 1997), p. 79; eprint quant-ph/9611006.

\bibitem{cleve98} H. Buhrman, R. Cleve, and A. Wigderson, ``Quantum
vs. Classical Communication and Computation,'' in {\em Proc. of the
30th Ann.  ACM Symp. on the Theory of Computing} (ACM Press, 1998),
p. 63; eprint quant-ph/9802040.

\bibitem{bbcjpw} C. H. Bennett, G. Brassard, C. Crepeau, R. Jozsa,
A. Peres, W. K. Wootters, ``Teleporting an Unknown Quantum State via
Dual Classical and Einstein-Podolsky-Rosen Channels,'' Phys.
Rev. Lett. {\bf 70}, 1895 (1993).

\bibitem{PS} P. W. Shor, ``Polynomial time algorithms for prime
factorization and discrete logarithms on a quantum computer,'' SIAM
J. Comput. {\bf26}, 1484 (1997), and references therein.

\bibitem{Grover} L. Grover, ``Quantum mechanics helps in searching for
a needle in a haystack,'' Phys. Rev. Lett. {\bf79}, 325 (1997).

\bibitem{WZ} W. K. Wootters and W. Zurek, Nature (London) {\bf299},
802 (1982).

\bibitem{PV} S. Popescu and L. Vaidman, ``Causality constraints on
nonlocal quantum measurements,'' Phys. Rev. A {\bf49}, 4331 (1994).

\bibitem{PW} A. Peres and W. K. Wootters, ``Optimal Detection of
Quantum Information,'' Phys. Rev. Lett. {\bf66}, 1119 (1991).

\bibitem{Holevo79}
A.~S. Kholevo, ``On the Capacity of Quantum Communication Channel,''
Prob.\ Inf.\ Transm.\ {\bf 15}, 247 (1979).

\bibitem{Holevo96} A.~S. Holevo, ``The Capacity of Quantum
Communication Channel with General Signal States,'' IEEE Trans.\ Inf.\
Theor. {\bf44}, 269 (1998); e-print quant-ph/9611023.

\bibitem{Schumacher97a}
B.~Schumacher and M.~D. Westmoreland,  ``Sending Classical
Information via Noisy Quantum Channels,'' Phys.\ Rev.\ A {\bf 56},
131 (1997).

\bibitem{Haus} P. Hausladen, R. Jozsa, B. Schumacher, M. Westmoreland,
and W. K. Wootters, ``Classical information capacity of a quantum
channel,'' Phys. Rev. A {\bf54}, 1869 (1996).

\bibitem{MP} S. Massar and S. Popescu, ``Optimal extraction of
information from finite quantum ensembles,'' Phys. Rev. Lett. {\bf74},
1259 (1995).

\bibitem{GV} L. Goldenberg and L. Vaidman, ``Quantum cryptography with
orthogonal states,'' Phys. Rev. Lett. {\bf 75}, 1239 (1995).

\bibitem{Asher} A. Peres, ``Quantum cryptography with orthogonal
states?''  Phys. Rev. Lett. {\bf 77}, 3264 (1996); L. Goldenberg and
L. Vaidman, ``Goldenberg and Vaidman Reply,'' {\em ibid.} {\bf 77},
3265 (1996).

\bibitem{Tal98} T. Mor, ``No-cloning of orthogonal states in composite
systems,'' Phys. Rev. Lett. {\bf80}, 3137 (1998); quant-ph/9802036.

\bibitem{BB84} C. H. Bennett and G. Brassard, ``Quantum Cryptography:
Public Key Distribution and Coin Tossing,'' in {\em Proceedings of the
IEEE International Conference on Computers, Systems and Signal
Processing, Bangalore, India} (IEEE, New York, 1984), p. 175.

\bibitem{Rains} E. Rains, ``Entanglement purification via separable
superoperators,'' quant-ph/9707002.

\bibitem{VP} V. Vedral and M. B. Plenio, ``Entanglement measures and
purification procedures,'' Phys. Rev. A {\bf57}, 1619 (1998),
quant-ph/9707035.

\bibitem{Horod} M. Horodecki, P. Horodecki, and R. Horodecki,
``Mixed-state entanglement and distillation: is there `bound'
entanglement in nature?''  quant-ph/9801069.

\bibitem{Bar} H. Barnum, M. A. Nielsen, and B. Schumacher,
``Information Transmission Through a Noisy Quantum Channel,''
quant-ph/9702049.

\bibitem{supops} B. Schumacher, ``Sending entanglement through noisy
quantum channels,'' Phys. Rev. A {\bf 54}, 2614 (1996);
quant-ph/9604023.

\bibitem{Niels} M. A. Nielsen, C. M. Caves, B. Schumacher, and
H. Barnum, ``Information-theoretic Approach to Quantum Error
Correction and Reversible Measurement,'' Proc. R. Soc. Lond. A {\bf
454}, 277 (1998); quant-ph/9706064.

\bibitem{Aharonov} Weak measurements were introduced by Y. Aharonov,
D. Z. Albert, and L. Vaidman, ``How the result of a measurement of a
component of the spin of a spin-1/2 particle can turn out to be 100,''
Phys. Rev. Lett. {\bf60}, 1351 (1988).  See also Y. Aharonov and
L. Vaidman, ``Properties of a quantum system during the time interval
between two measurements,'' Phys. Rev. A {\bf41}, 11 (1990), Sec. IV.
The quantum gates for our weak measurements (used in our proof in
Appendix \protect\ref{peter}) were introduced by T. Mor, PhD Thesis,
Sec. 4.4 (unpublished).

\bibitem{Abr} N. Abramson, {\em Information Theory and Coding}
(McGraw-Hill, New York, 1963), Chap. 5.

\bibitem{OhyaPetz}
M.~Ohya and D.~Petz, {\em Quantum Entropy and Its Use}
(Springer-Verlag, Berlin, 1993).

\bibitem{LevitinFuchs} For distinguishing two equiprobable pure states
with overlap $\delta=\cos(2\theta)$, the optimal measurement is a von
Neumann measurement in the subspace spanned by the two states, in a
basis symmetrically disposed around these two states.  The resulting
error probability is $p={1\over 2}(1-\sin(2\theta))$, so that the last
two terms on the right-hand side of Eq. (\protect\ref{DisBeLloydBaby})
become $1-h(p)$.  See L.~B. Levitin, ``Optimal quantum measurements
for two pure and mixed states,'' in {\em Quantum Communications and
Measurement\/} (eds. V.~P. Belavkin, O.~Hirota, and R.~L. Hudson,
Plenum Press, New York, 1995), and C.~A. Fuchs and C.~M. Caves,
``Ensemble-dependent bounds for accessible information in quantum
mechanics,'' Phys. Rev. Lett. {\bf 73}, 3047 (1994).

\bibitem{BS} B. Schumacher, ``Quantum Coding,'' Phys. Rev. A {\bf51},
2738 (1995).

\bibitem{BDSW} C.~H. Bennett, D.~P. DiVincenzo, J.~A. Smolin,
W.~K. Wootters, ``Mixed state entanglement and quantum
error-correction,'' Phys. Rev. A {\bf54}, 3824 (1996); e-print
quant-ph/9604024.

\bibitem{Landauer} R.~Landauer, ``Irreversibility and Heat Generation
in the Computing Process,'' IBM J. Res. Develop.  {\bf 5},
183-191 (1961).

\bibitem{foot4} It is easy to show by a process of elimination that
there does not exist any product vector which is orthogonal to the
four states of Eq. (\protect\ref{4states}).  From this we find that
the density operator ${1\over 4}(I-\sum_{k=1}^4 P_k)$ has many
fascinating entanglement properties.  Among other things, it exhibits
the ``bound'' entanglement of Ref. \protect\cite{Horod}.  This will be
explored in a forthcoming publication (D.~P. DiVincenzo, T. Mor, P.~W.
Shor, J.~A. Smolin, and B.~M. Terhal, ``Unextendible product bases
and bound entanglement,'' quant-ph/9808030.

\bibitem{Keller} J. C. Lagarias and P. W. Shor, ``Keller's cube-tiling
conjecture is false in higher dimensions,'' Bull Amer. Math. Soc. {\bf
27}, 279 (1992)
(see http://www.research.att.com/\~\,shor/papers/index.html);
``Cube tilings of $R^N$ and nonlinear
codes,'' Discrete Comput. Geom. {\bf11}, 359 (1994).

\bibitem{Pop} In an earlier version of this paper a conjecture by
Popescu, that the states $\cos\theta\ket{00}+\sin\theta\ket{11}$ and
$-\sin\theta\ket{00}+\cos\theta\ket{11}$ might not be locally
distinguishable given any finite number of copies, has since been
shown to be false.  In fact these states can be distinguished given
only one copy of the state (private discussions with S. Popescu and
N. Linden, 1998).

\bibitem{Hoy} P. H{\o}yer, private communication.

\bibitem{BBM} C. H. Bennett, G. Brassard, and N. D. Mermin, ``Quantum
cryptography without Bell's theorem,'' Phys. Rev. Lett. {\bf68}, 557
(1992).

\bibitem{butsee} But, see D. Aharonov, A. Kitaev, and N. Nisan,
``Quantum Circuits with Mixed States,'' Proc. 13th Annu. Symp. on the
Theory of Computation (ACM Press, Los Alamitos, 1998), p. 20
(quant-ph/9806029); S. Richter and R.~F. Werner, ``Ergodicity of
quantum cellular automata,'' J. Stat. Phys. {\bf82}, 963 (1996)
(cond-mat/9504001).

\bibitem{Fuchs96a} C.~A. Fuchs and A.~Peres, ``Quantum State
Disturbance vs.\ Information Gain: Uncertainty Relations for Quantum
Information,'' Phys.\ Rev.\ A {\bf 53}, 2038 (1996).

\bibitem{Caves} C. M. Caves, K. S. Thorne, R. W. P. Drever,
V. D. Sandberg, and M. Zimmermann, ``On the measurement of a weak
classcial force coupled to a quantum-mechanical oscillator. I. Issues
of principle,'' Rev. Mod. Phys. {\bf 52}, 341 (1980).

\bibitem{Ber} E. Kranakis, {\em Primality and Cryptography}
(J. Wiley, New York, 1986), p. 94; A. R\'enyi, {\em Foundations of
Probability} (Holden-Day, San Francisco, 1970), p. 200.

\bibitem{Weg} M. N. Wegman and J. L. Carter, J. of Computer and System
Sciences {\bf22}, 265 (1981).

\end{references}
\end{document}